\documentclass[11pt,a4paper]{article}
\pdfoutput=1
\usepackage{jheppub}
\usepackage{lscape}
\usepackage{subfigure}
\usepackage{multirow}
\usepackage{lineno}

\long\def\symbolfootnote[#1]#2{\begingroup%
\def\thefootnote{\fnsymbol{footnote}}\footnote[#1]{#2}\endgroup}

\def\hermesauthor[#1]#2{{#2}$^{\, #1}$}
\def\hermesinstitute[#1]#2{$^{#1\,}$ {#2}\\}

\renewcommand{\thefootnote}{\alph{footnote}}
\def\nowat[#1]#2{\(^,\)\footnote[#1]{#2}}

\def\desy{{\sc Desy}}

\title{Beam-helicity and beam-charge asymmetries 
	associated with deeply virtual Compton scattering
         on the unpolarised proton}

\author{The {\sc Hermes} Collaboration}

\affiliation{DESY -- HERMES, Notkestra\ss e 85, D-22607 Hamburg}
\emailAdd{management@hermes.desy.de}

\abstract{
Beam-helicity and beam-charge asymmetries in the hard exclusive leptoproduction of real photons from an unpolarised hydrogen target by a 27.6\,GeV lepton beam 
are extracted from the H{\sc ermes} data set of 2006-2007 using a
missing-mass event selection technique. The asymmetry
amplitudes extracted from this data set are more
precise than those extracted from the earlier data
set of 1996-2005 previously analysed in the same manner by H{\sc ermes}. The results from the two data sets are 
compatible with each other. Results from these combined data sets are
extracted and constitute the most precise asymmetry amplitude measurements made in the H{\sc ermes} kinematic region using a missing-mass event selection technique.
}
\keywords{lepton-nucleon scattering}
\notoc

\begin{document}

\maketitle\flushbottom\newpage

\pagestyle{myplain}
\pagenumbering{roman}

                                    
\section*{Author list}
\subsection*{The {\sc Hermes} Collaboration}

{%
\begin{flushleft} 
\bf
\hermesauthor[13,16]{A.~Airapetian},
\hermesauthor[27]{N.~Akopov},
\hermesauthor[6]{Z.~Akopov},
\hermesauthor[7]{E.C.~Aschenauer}\nowat[1]{Now at: Brookhaven National Laboratory, Upton, New York 11772-5000, USA},
\hermesauthor[26]{W.~Augustyniak},
\hermesauthor[27]{R.~Avakian},
\hermesauthor[27]{A.~Avetissian},
\hermesauthor[6]{E.~Avetisyan},
\hermesauthor[18,25]{H.P.~Blok},
\hermesauthor[6]{A.~Borissov},
\hermesauthor[14]{J.~Bowles},
\hermesauthor[20]{V.~Bryzgalov},
\hermesauthor[14]{J.~Burns},
\hermesauthor[10]{M.~Capiluppi},
\hermesauthor[11]{G.P.~Capitani},
\hermesauthor[22]{E.~Cisbani},
\hermesauthor[10]{G.~Ciullo},
\hermesauthor[10]{M.~Contalbrigo},
\hermesauthor[10]{P.F.~Dalpiaz},
\hermesauthor[6]{W.~Deconinck},
\hermesauthor[2]{R.~De~Leo},
\hermesauthor[12,6,23]{L.~De~Nardo},
\hermesauthor[11]{E.~De~Sanctis},
\hermesauthor[15,9]{M.~Diefenthaler},
\hermesauthor[11]{P.~Di~Nezza},
\hermesauthor[13]{M.~D\"uren},
\hermesauthor[13]{M.~Ehrenfried},
\hermesauthor[27]{G.~Elbakian},
\hermesauthor[5]{F.~Ellinghaus},
\hermesauthor[7]{R.~Fabbri},
\hermesauthor[11]{A.~Fantoni},
\hermesauthor[23]{L.~Felawka},
\hermesauthor[22]{S.~Frullani},
\hermesauthor[7]{D.~Gabbert},
\hermesauthor[20]{G.~Gapienko},
\hermesauthor[20]{V.~Gapienko},
\hermesauthor[22]{F.~Garibaldi},
\hermesauthor[6,19,23]{G.~Gavrilov},
\hermesauthor[15,10]{F.~Giordano},
\hermesauthor[16]{S.~Gliske},
\hermesauthor[7]{M.~Golembiovskaya},
\hermesauthor[11]{C.~Hadjidakis},
\hermesauthor[6]{M.~Hartig},
\hermesauthor[11]{D.~Hasch},
\hermesauthor[14]{M.~Hoek},
\hermesauthor[6]{Y.~Holler},
\hermesauthor[24]{Y.~Imazu},
\hermesauthor[1]{H.E.~Jackson},
\hermesauthor[12]{H.S.~Jo},
\hermesauthor[14]{R.~Kaiser}\nowat[2]{Present address: International Atomic Energy Agency, A-1400 Vienna, Austria},
\hermesauthor[27]{G.~Karyan},
\hermesauthor[14,13]{T.~Keri},
\hermesauthor[5]{E.~Kinney},
\hermesauthor[19]{A.~Kisselev},
\hermesauthor[24]{N.~Kobayashi},
\hermesauthor[20]{V.~Korotkov},
\hermesauthor[17]{V.~Kozlov},
\hermesauthor[9,19]{P.~Kravchenko},
\hermesauthor[8]{V.G.~Krivokhijine},
\hermesauthor[2]{L.~Lagamba},
\hermesauthor[18]{L.~Lapik\'as},
\hermesauthor[14]{I.~Lehmann},
\hermesauthor[10]{P.~Lenisa},
\hermesauthor[16]{W.~Lorenzon},
\hermesauthor[3]{B.-Q.~Ma},
\hermesauthor[14]{D.~Mahon},
\hermesauthor[15]{N.C.R.~Makins},
\hermesauthor[19]{S.I.~Manaenkov},
\hermesauthor[22]{L.~Manfr\'e},
\hermesauthor[3]{Y.~Mao},
\hermesauthor[26]{B.~Marianski},
\hermesauthor[6,5]{A.~Martinez de la Ossa},
\hermesauthor[27]{H.~Marukyan},
\hermesauthor[23]{C.A.~Miller},
\hermesauthor[24]{Y.~Miyachi}\nowat[3]{Now at: Department of Physics, Yamagata University, Yamagata 990-8560, Japan},
\hermesauthor[27]{A.~Movsisyan},
\hermesauthor[11]{V.~Muccifora},
\hermesauthor[14]{M.~Murray},
\hermesauthor[6,9]{A.~Mussgiller},
\hermesauthor[2]{E.~Nappi},
\hermesauthor[19]{Y.~Naryshkin},
\hermesauthor[9]{A.~Nass},
\hermesauthor[7]{W.-D.~Nowak},
\hermesauthor[10]{L.L.~Pappalardo},
\hermesauthor[13]{R.~Perez-Benito},
\hermesauthor[27]{A.~Petrosyan},
\hermesauthor[9]{M.~Raithel},
\hermesauthor[1]{P.E.~Reimer},
\hermesauthor[11]{A.R.~Reolon},
\hermesauthor[7]{C.~Riedl},
\hermesauthor[9]{K.~Rith},
\hermesauthor[14]{G.~Rosner},
\hermesauthor[6]{A.~Rostomyan},
\hermesauthor[1,15]{J.~Rubin},
\hermesauthor[12]{D.~Ryckbosch},
\hermesauthor[20]{Y.~Salomatin},
\hermesauthor[24,21]{F.~Sanftl},
\hermesauthor[21]{A.~Sch\"afer},
\hermesauthor[4,12]{G.~Schnell},
\hermesauthor[6]{K.P.~Sch\"uler},
\hermesauthor[14]{B.~Seitz},
\hermesauthor[24]{T.-A.~Shibata},
\hermesauthor[8]{V.~Shutov},
\hermesauthor[10]{M.~Stancari},
\hermesauthor[10]{M.~Statera},
\hermesauthor[9]{E.~Steffens},
\hermesauthor[18]{J.J.M.~Steijger},
\hermesauthor[7]{J.~Stewart},
\hermesauthor[27]{S.~Taroian},
\hermesauthor[17]{A.~Terkulov},
\hermesauthor[15]{R.~Truty},
\hermesauthor[26]{A.~Trzcinski},
\hermesauthor[12]{M.~Tytgat},
\hermesauthor[12]{Y.~Van~Haarlem},
\hermesauthor[4,12]{C.~Van~Hulse}, 
\hermesauthor[19]{D.~Veretennikov},
\hermesauthor[19]{V.~Vikhrov},
\hermesauthor[2]{I.~Vilardi},
\hermesauthor[3]{S.~Wang},
\hermesauthor[7,9]{S.~Yaschenko},
\hermesauthor[6]{Z.~Ye},
\hermesauthor[23]{S.~Yen},
\hermesauthor[13]{W.~Yu},
\hermesauthor[6,16]{V.~Zagrebelnyy},
\hermesauthor[9]{D.~Zeiler},
\hermesauthor[6]{B.~Zihlmann},
\hermesauthor[26]{P.~Zupranski}
\end{flushleft} 
}
\bigskip
{\it
\begin{flushleft} 
\hermesinstitute[1]{Physics Division, Argonne National Laboratory, Argonne, Illinois 60439-4843, USA}
\hermesinstitute[2]{Istituto Nazionale di Fisica Nucleare, Sezione di Bari, 70124 Bari, Italy}
\hermesinstitute[3]{School of Physics, Peking University, Beijing 100871, China}
\hermesinstitute[4]{Department of Theoretical Physics, University of the Basque Country UPV/EHU, 48080 Bilbao, Spain and IKERBASQUE, Basque Foundation for Science, 48011 Bilbao, Spain}
\hermesinstitute[5]{Nuclear Physics Laboratory, University of Colorado, Boulder, Colorado 80309-0390, USA}
\hermesinstitute[6]{DESY, 22603 Hamburg, Germany}
\hermesinstitute[7]{DESY, 15738 Zeuthen, Germany}
\hermesinstitute[8]{Joint Institute for Nuclear Research, 141980 Dubna, Russia}
\hermesinstitute[9]{Physikalisches Institut, Universit\"at Erlangen-N\"urnberg, 91058 Erlangen, Germany}
\hermesinstitute[10]{Istituto Nazionale di Fisica Nucleare, Sezione di Ferrara and Dipartimento di Fisica, Universit\`a di Ferrara, 44100 Ferrara, Italy}
\hermesinstitute[11]{Istituto Nazionale di Fisica Nucleare, Laboratori Nazionali di Frascati, 00044 Frascati, Italy}
\hermesinstitute[12]{Department of Physics and Astronomy, Ghent University, 9000 Gent, Belgium}
\hermesinstitute[13]{Physikalisches Institut, Universit\"at Gie{\ss}en, 35392 Gie{\ss}en, Germany}
\hermesinstitute[14]{SUPA, School of Physics and Astronomy, University of Glasgow, Glasgow G12 8QQ, United Kingdom}
\hermesinstitute[15]{Department of Physics, University of Illinois, Urbana, Illinois 61801-3080, USA}
\hermesinstitute[16]{Randall Laboratory of Physics, University of Michigan, Ann Arbor, Michigan 48109-1040, USA }
\hermesinstitute[17]{Lebedev Physical Institute, 117924 Moscow, Russia}
\hermesinstitute[18]{National Institute for Subatomic Physics (Nikhef), 1009 DB Amsterdam, The Netherlands}
\hermesinstitute[19]{B.P. Konstantinov Petersburg Nuclear Physics Institute, Gatchina, 188300 Leningrad Region, Russia}
\hermesinstitute[20]{Institute for High Energy Physics, Protvino, 142281 Moscow Region, Russia}
\hermesinstitute[21]{Institut f\"ur Theoretische Physik, Universit\"at Regensburg, 93040 Regensburg, Germany}
\hermesinstitute[22]{Istituto Nazionale di Fisica Nucleare, Sezione di Roma, Gruppo Collegato Sanit\`a and Istituto Superiore di Sanit\`a, 00161 Roma, Italy}
\hermesinstitute[23]{TRIUMF, Vancouver, British Columbia V6T 2A3, Canada}
\hermesinstitute[24]{Department of Physics, Tokyo Institute of Technology, Tokyo 152, Japan}
\hermesinstitute[25]{Department of Physics and Astronomy, VU University, 1081 HV Amsterdam, The Netherlands}
\hermesinstitute[26]{National Centre for Nuclear Research, 00-689 Warsaw, Poland}
\hermesinstitute[27]{Yerevan Physics Institute, 375036 Yerevan, Armenia}
\end{flushleft} 
}

\clearpage
\pagenumbering{arabic}

\noindent\rule\textwidth{.1pt}
\tableofcontents
\afterTocSpace
\noindent\rule\textwidth{.1pt}

\pagestyle{myplain}

\section{Introduction}

\begin{figure}
\centering
\subfigure[DVCS]{\includegraphics[height=0.34\textwidth]{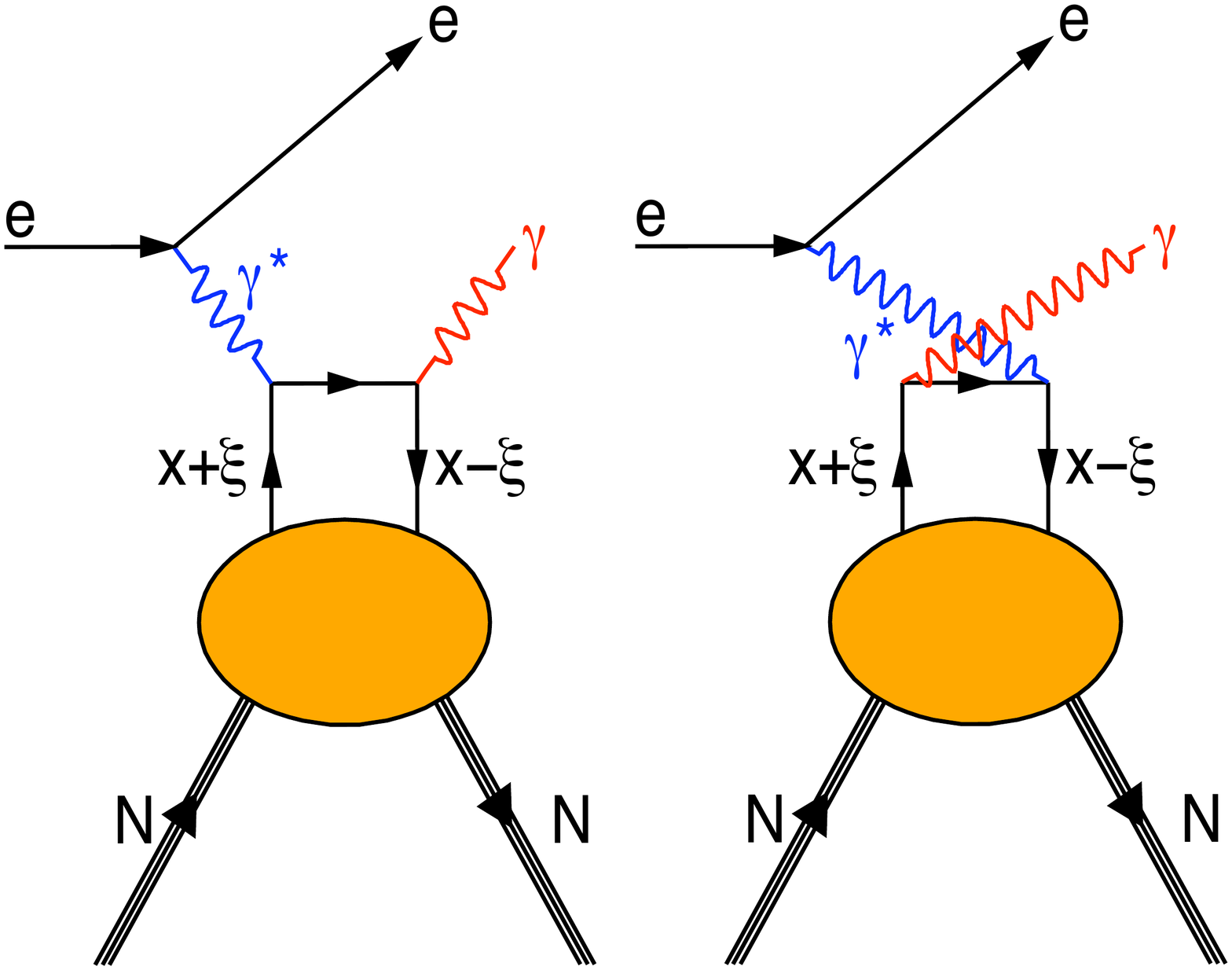}}
\hspace*{1,2cm}\subfigure[Bethe-Heitler]{\includegraphics[height=0.34\textwidth]{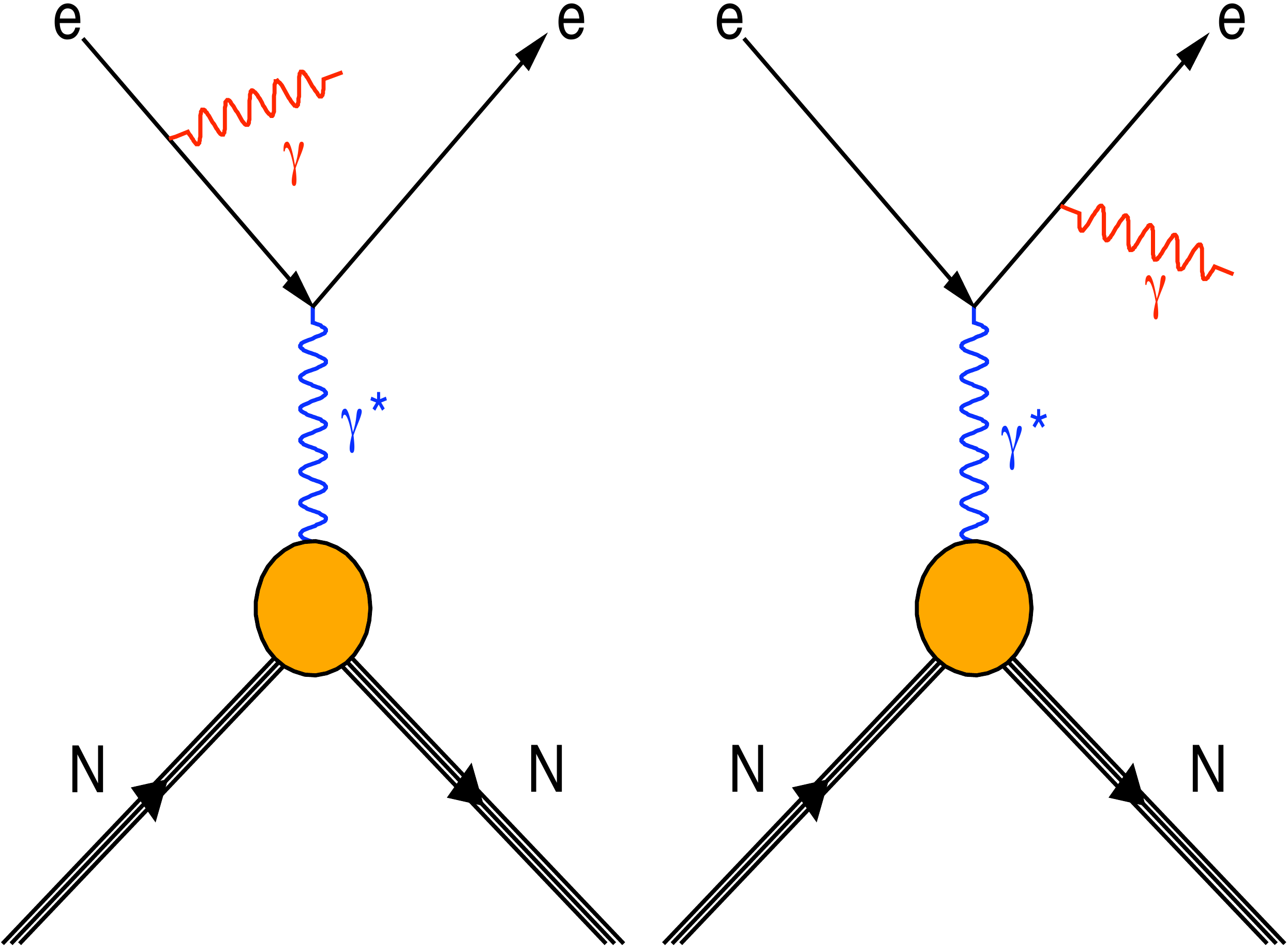}}
\caption[DVCS and Bethe Heitler hand bag diagram.]{(a): The leading DVCS process in which an electron/positron ($e$) interacts with a quark in the nucleon
($N$) via a virtual photon ($\gamma^\ast$). The quark is found in the
nucleon with longitudinal momentum fraction $x+\xi$ and emits a real
photon ($\gamma$). The quark is absorbed by the nucleon with
longitudinal momentum fraction $x-\xi$. (b): The leading Bethe-Heitler process, i.e. the emission of a real photon from the incoming or outgoing lepton. This process has the same initial and final states as DVCS.}
\label{spin}
\end{figure}

Generalised Parton Distributions (GPDs)
\cite{Mue94,Ji97,Rad97} encompass the familiar Parton
Distribution Functions (PDFs) and nucleon Form Factors (FFs) to provide a
comprehensive description of the structure of the nucleon. A thorough
description of the nucleon in terms of GPDs would allow the deduction
of the total angular momentum of partons in the nucleon, and the
construction of a longitudinal-momentum-dissected transverse spatial map of parton densities~\cite{Bur00}. The GPDs appear in experimental measurements in the form of complex-valued Compton Form Factors (CFFs), which are flavour-sums of convolutions of GPDs with hard scattering kernels. Constraints on these CFFs, and thus GPDs, can be obtained from measurements of exclusive leptoproduction processes. In particular, the exclusive leptoproduction of a single real photon from a nucleon or nucleus that remains intact ($e\,N\,\rightarrow\,e\,N\,\gamma$; see figure~\ref{spin}) is the simplest to describe and is the most widely-used reaction channel for such work (see \cite{Ji97b,Air01,Air06,Air08,Air09,Air10,Air10a,Air10b,Air11,Air11a,h101,h105,h107,h109,zeu03,zeu08,Maz07,Cam06,Ste01,Che06,Gir08,Gav09,Com10}).

Generalised parton distributions depend upon four kinematic variables: the
Mandelstam variable $t=(p-p^{\prime})^2$, which is the squared momentum
transfer to the target nucleon in the exclusive scattering process with $p$ ($p^{\prime}$)
representing the initial (final) four-momentum of the nucleon; the average
fraction $x$ of the nucleon's longitudinal momentum carried by the active
quark throughout the scattering process; half the difference of
  the fractions of the nucleon's longitudinal momentum carried
by the active quark at the start and end of the process, written as
the skewness $\xi$; and $Q^2=-(q^2)$, i.e. the negative square of the four-momentum of
the virtual photon that mediates the lepton-nucleon scattering
process. In the Bjorken limit of $Q^2\rightarrow\infty$ with fixed
$t$, the skewness $\xi$ is related to the Bjorken variable
$x_{\textrm{B}}=\frac{-q^2}{2p\cdot q}$ as
$\xi\approx\frac{x_\textrm{B}}{2-x_\textrm{B}}$. The results are presented
as a function of $x_{\textrm{B}}$ because there is no consensus on an experimentally observable representation of $\xi$. 
Exclusive leptoproduction of real photons
 arises from
two experimentally indistinguishable processes: the Deeply Virtual Compton Scattering (DVCS) process,
which is the emission of a real photon by the struck quark from the nucleon, and the Bethe--Heitler (BH) process, which is elastic lepton-nucleon scattering with the emission of a bremsstrahlung photon by the lepton.
The BH process is calculable in the QED framework; this process is
dominant at the kinematic conditions of the H{\sc ermes} experiment. The
two processes interfere and the large BH amplitude
amplifies the interference term, which is proportional to the DVCS amplitude. It is through the study of this interference term that useful information for the constraint of certain GPDs can be obtained at H{\sc ermes} kinematic conditions, especially since the interference term is the only part of the squared scattering amplitude that is linear in CFFs~\cite{Bel02b}.

The four-fold differential cross section for the exclusive leptoproduction of real photons
from an unpolarised hydrogen target can be written as \cite{Bel02b}
\begin{equation}
\frac{\textrm{d}^4\sigma}{\textrm{d}x_{\textrm{B}}\textrm{d}Q^{2}\textrm{d}
|t|\textrm{d}\phi} =
\frac{x_{\textrm{B}}e^{6}}{32(2\pi)^{4} Q^{4}\sqrt{1+\epsilon^{2}}}
|\tau|^{2},
\end{equation}
where $e$ is the elementary
charge, $\epsilon=2x_\textrm{B}\frac{M}{Q}$ with $M$
the target mass, and $\phi$ is the
azimuthal angle between the scattering and production planes \cite{Tre04}.
The {square of the} scattering amplitude $|\tau|^2$ can be written as
\begin{equation}
|\tau|^{2} = |\tau_{\textrm{BH}}|^{2} +
|\tau_{\textrm{DVCS}}|^{2} + \textrm{I},
\end{equation}
with contributions from the \textrm{BH} process ($\lvert\tau_{\textrm{BH}}\rvert^2$),
the DVCS process
($\lvert\tau_{\textrm{DVCS}}\rvert^2$) and their interference term (I). These
contributions can be written as
\begin{eqnarray}
 |\tau_{\textrm{BH}}|^{2} &=&
 \frac{K_{\textrm{BH}}}{\mathcal{P}_{1}(\phi)\mathcal{P}_{2}(\phi)} \left(c_{\textrm{unp},0}^{\textrm{BH}} + \sum_{n=1}^2
  c_{\textrm{unp},n}^{\textrm{BH}}\cos(n\phi)\right), \label{e:tbh}\\
|\tau_{\textrm{DVCS}}|^{2} &=&
K_{\textrm{DVCS}}\left(c_{\textrm{unp},0}^{\textrm{DVCS}} +
\sum_{n=1}^2
c_{\textrm{unp},n}^{\textrm{DVCS}}\cos(n\phi) + \lambda\,
s_{\textrm{unp},1}^{\textrm{DVCS}}\sin\phi\right)\,\textrm{and}
\label{e:tdvcs}\\
 \textrm{I} &=& \frac{- e_\ell
K_{\textrm{I}}}{\mathcal{P}_{1}(\phi)\mathcal{P}_{2}(\phi)}\left(c_{\textrm{unp},0}^{\textrm{
I}}+
\sum_{n=1}^3 c_{\textrm{unp},n}^{\textrm{I}}\cos(n\phi) + \lambda \sum_{n=1}^2
s_{\textrm{unp},n}^{\textrm{I}}\sin(n\phi)\right),\label{e:ti}
\end{eqnarray}
where $\mathcal{P}_1(\phi)$ and $\mathcal{P}_2(\phi)$ are the lepton propagators
of the BH process, $\lambda$ is the
helicity of the lepton beam and $e_\ell$ is the sign of the charge of
 the beam lepton.  The
quantities $K_{\textrm{BH}}=1/(x_\textrm{B}^2t(1+\epsilon^2)^2)$,
$K_{\textrm{DVCS}}=1/Q^2$
and $K_{\textrm{I}}=1/(x_{\textrm{B}}yt)$ are kinematic factors, where
$y$ is the fraction of the beam energy carried by the virtual photon in
the target rest frame. A full explanation of the Fourier coefficients [$c_{\textrm{unp},n}^V,s_{\textrm{unp},n}^W$], where V (W) denotes BH, DVCS or I (DVCS or I), can be found in ref.~\cite{Bel02b}.
 
Two sets of asymmetries measured at H{\sc
ermes} with an unpolarised hydrogen target and a polarised electron or positron
beam are considered here:
beam-helicity asymmetries and beam-charge asymmetries. This paper,
like ref.~\cite{Air09}, presents results related to the following asymmetries:
\begin{align}
\hspace{0.5cm}\mathcal{A}^{\textrm{I}}_{\textrm{LU}}(\phi) &\equiv
\frac{(\textrm{d}\sigma(\phi)^{+\rightarrow} -
\textrm{d}\sigma(\phi)^{+\leftarrow}) -
(\textrm{d}\sigma(\phi)^{-\rightarrow}
- \textrm{d}\sigma(\phi)^{-\leftarrow})}{(\textrm{d}\sigma(\phi)^{+\rightarrow}
+
\textrm{d}\sigma(\phi)^{+\leftarrow}) +
(\textrm{d}\sigma(\phi)^{-\rightarrow}
+ \textrm{d}\sigma(\phi)^{-\leftarrow})}&  \nonumber \\
&=\dfrac{-\dfrac{K_{\textrm{I}}}{\mathcal{P}_{1}(\phi)\mathcal{P}_{2}(\phi)}
\displaystyle\sum_{n=1}^2
s_{\textrm{unp},n}^{\textrm{I}}\sin(n\phi)}{\dfrac{K_{\textrm{BH}}}{\mathcal{P}_{1}
(\phi)\mathcal{P}_{
2}(\phi)}
\displaystyle\sum_{n=0}^2
c_{\textrm{unp},n}^{\textrm{BH}}\cos(n\phi) + 
K_{\textrm{DVCS}}\displaystyle\sum_{n=0}^2 c_{\textrm{unp},n}^{\textrm{DVCS}}\cos(n\phi)},& 
\label{e:alui}
\end{align}

\begin{align}
\mathcal{A}^{\textrm{DVCS}}_{\textrm{LU}}(\phi) &\equiv
\frac{(\textrm{d}\sigma(\phi)^{+\rightarrow} +
\textrm{d}\sigma(\phi)^{-\rightarrow}) -
(\textrm{d}\sigma(\phi)^{+\leftarrow} + 
\textrm{d}\sigma(\phi)^{-\leftarrow})}
{(\textrm{d}\sigma(\phi)^{+\rightarrow} +
\textrm{d}\sigma(\phi)^{-\rightarrow}) +
(\textrm{d}\sigma(\phi)^{+\leftarrow}
+ \textrm{d}\sigma(\phi)^{-\leftarrow})}& \nonumber \\[0.4cm]
&=\dfrac{ K_{\textrm{DVCS}} s_{\textrm{unp},1}^{\textrm{DVCS}}\sin\phi}{\dfrac{K_{\textrm{BH}}}{\mathcal{P}_{1}
(\phi)\mathcal{P}_{2}(\phi)}
\displaystyle\sum_{n=0}^2
c_{\textrm{unp},n}^{\textrm{BH}}\cos(n\phi) + 
K_{\textrm{DVCS}}\displaystyle\sum_{n=0}^2
c_{\textrm{unp},n}^{\textrm{DVCS}}\cos(n\phi)}\, , &
\label{e:aludvcs}
\end{align}

\begin{align}
\hspace{0.5cm}\mathcal{A}_{\textrm{C}}(\phi) &\equiv  
\frac{(\textrm{d}\sigma(\phi)^{+\rightarrow} +
\textrm{d}\sigma(\phi)^{+\leftarrow}) -
(\textrm{d}\sigma(\phi)^{-\rightarrow}
+ \textrm{d}\sigma(\phi)^{-\leftarrow})}{(\textrm{d}\sigma(\phi)^{+\rightarrow}
+
\textrm{d}\sigma(\phi)^{+\leftarrow}) +
(\textrm{d}\sigma(\phi)^{-\rightarrow}
+ \textrm{d}\sigma(\phi)^{-\leftarrow})}&    \nonumber \\
&=\dfrac{{-\dfrac{K_{\textrm{I}}}{\mathcal{P}_{1}(\phi)\mathcal{P}_{2}(\phi)}
\displaystyle\sum_{n=0}^3
c_{\textrm{unp},n}^{\textrm{I}}\cos(n\phi)}}{\dfrac{K_{\textrm{BH}}}{\mathcal{P}_{1}
(\phi)\mathcal{P}_
{2}(\phi)}
\displaystyle\sum_{n=0}^2
c_{\textrm{unp},n}^{\textrm{BH}}\cos(n\phi) + 
K_{\textrm{DVCS}}\displaystyle\sum_{n=0}^2 c_{\textrm{unp},n}^{\textrm{DVCS}}\cos(n\phi)} ,&
\label{e:ac}
\end{align}
where $\textrm{d}\sigma(\phi)^+$ ($\textrm{d}\sigma(\phi)^-$) refers to
the differential cross section with positive (negative) beam charge and
$\textrm{d}\sigma(\phi)^\rightarrow$ ($\textrm{d}\sigma(\phi)^\leftarrow$) refers
to the differential cross section taken with beam spin parallel (anti-parallel) to the
beam momentum.

The $s_{\textrm{unp},n}^{W}$ and $c_{\textrm{unp},n}^{W}$ Fourier
coefficients depend on ``$\mathcal{C}$-functions'' \cite{Bel02b}, each of which is a
combination of CFFs. Contributions to the cross section are suppressed by factors that may be kinematic in nature or due to the twist-level of the GPDs appearing in that contribution. Leading twist is twist-2. Typically, the contribution of a twist-$n$ GPD, and hence the corresponding CFF, is
suppressed by $\mathcal{O}(1/Q^{n-2})$. 

The Fourier coefficients that receive leading-twist contributions are $c_{\textrm{unp},0}^{\textrm{I}}$, $c_{\textrm{unp},1}^{\textrm{I}}$ and $s_{\textrm{unp},1}^{\textrm{I}}$. All of these Fourier coefficients have a dominant contribution from the $\mathcal{C}_{\textrm{unp}}^{\textrm{I}}$-function:
\begin{eqnarray}
c_{\textrm{unp},0}^{\textrm{I}} &\approx&-8(2-y)\frac{(2-y^2)}{(1-y)}K^2\mathfrak{Re}\mathcal{C}_{\textrm{unp}}^{\textrm{I}}\label{eq:c0}
\\
c_{\textrm{unp},1}^{\textrm{I}} &\approx&\,8K(2- 2y + y^{2})\mathfrak{Re}\mathcal{C}_{\textrm{unp}}^{\textrm{I}}\,,\label{eq:c1}
\\
s_{\textrm{unp},1}^{\textrm{I}} &\approx&\,8Ky(2-y)\mathfrak{Im}\mathcal{C}_{\textrm{unp}}^{\textrm{I}}\,. \label{eq:s1}
\end{eqnarray}
The definition of the kinematic factor $K$ is~\cite{Bel02b}:
\begin{equation}
K^2=\frac{t}{Q^2}\Big(1-\frac{t_{\textrm{min}}}{t}\Big)\Big(1-x_{\textrm{B}}\Big)\Big(1-y-\frac{y^2\epsilon^2}{4}\Big)\Big\lbrace\sqrt{1+\epsilon^2}+\frac{4x_{\textrm{B}}(1-x_{\textrm{B}})+\epsilon^2}{4(1-x_{\textrm{B}})}
\frac{t_{\textrm{min}}-t}{Q^2}\Big\rbrace\,.\label{eq:K}
\end{equation} 
The factor of $\Big(1-\frac{t_{\textrm{min}}}{t}\Big)$ implies that amplitudes proportional to these Fourier coefficients vanish as $-t$ approaches its minimum value.
The $\mathcal{C}_{\textrm{unp}}^{\textrm{I}}$-function can be
written
\cite{Bel02b} 
\begin{equation}
 \mathcal{C}_{\textrm{unp}}^{\textrm{I}} = F_{1}\mathcal{H} + \frac{x_{\textrm{B}}}{2-x_{\textrm{B}}}(F_{1}+F_{2})\widetilde{\mathcal{H}} -\frac{t}{4M^{2}}F_{2}\mathcal{E},
\label{Eq_cunp}
\end{equation}
where $F_{1}$ and $F_{2}$ are respectively the Dirac and Pauli form
factors of the nucleon and $\mathcal{H}$, $\widetilde{\mathcal{H}}$ and
$\mathcal{E}$ are CFFs that relate respectively to the GPDs $H$,
$\widetilde{H}$ and $E$.  In H{\sc ermes} kinematic
conditions (where $x_{\textrm{B}}$ and $\frac{-t}{4M^2}$ are of order 0.1), the
contributions of CFFs $\widetilde{\mathcal{H}}$ and $\mathcal{E}$ can be
neglected in eq.~\ref{Eq_cunp} with respect to $\mathcal{H}$ (in first approximation) since they
are kinematically suppressed by an order of magnitude or more.
Hence, the behaviour of
$\mathcal{C}_{\textrm{unp}}^{\textrm{I}}$ is determined by CFF $\mathcal{H}$
and therefore GPD $H$ can be constrained through
measurements of the $\sin\phi$ and $\cos\phi$ terms of the $\mathcal{A}^{\textrm{I}}_{\textrm{LU}}(\phi)$ and $\mathcal{A}_{\textrm{C}}(\phi)$ asymmetries respectively.

Compared to the analysis in ref.~\cite{Air09}, the analysis presented here additionally includes a larger, independent data set taken during the years 2006 and 2007 and makes use of the same missing-mass technique for event selection as was used in ref.~\cite{Air09}. The work covered in this publication further combines the data taken in 1996-2005 with this newer data set to produce the statistically most precise DVCS measurements that will be presented by H{\sc ermes}.

\section{Experiment and data selection}

The new data presented in this work were collected in 2006 and 2007. As in ref.~\cite{Air09}, the data were collected with the H{\sc ermes}
spectrometer \cite{Ack98} using the longitudinally polarised 27.6\,GeV
electron and positron beams incident upon an unpolarised hydrogen gas
target internal to the H{\sc era} lepton storage ring at D{\sc esy}. The integrated luminosities of the electron and positron data samples are
approximately 246\,pb$^{-1}$ and 1460\,pb$^{-1}$, with average beam polarisations of $0.303$ and $0.392$ respectively. The procedure used to select events is similar to that used in ref.~\cite{Air09}. A brief summary of this procedure is outlined in the following; more details are given in refs.~\cite{Zei09,Bur10}.

Events in the 2006-2007 data set were selected if having exactly one lepton
track detected within the acceptance of the spectrometer and
exactly one photon depositing $>5$ \,GeV in the electromagnetic
calorimeter. This photon is taken to be the photon arising from
  the process under investigation.
The latter selection criterion differs from the photon
selection criterion used for the 1996-2005 data set as an intermittent
hardware fault in 2006-2007 can cause spurious noise signals in the
calorimeter that are misinterpreted as very low energy photons.
The event selection is subject to the kinematic constraints 1\,GeV$^{2}$ $<$
Q$^{2}$ $<$ 10\,GeV$^{2}$, 0.03 $<$ $x_{\textrm{B}}$ $<$ 0.35,
$-t < 0.7$\,GeV$^2$, $W^{2}$ $>$
9\,GeV$^{2}$ and $\nu$ $<$ 22\,GeV, where $W$ is the invariant mass of the
$\gamma^{*}p$ system and $\nu$ is the energy of the virtual photon in the target
rest frame. The polar angle between the directions of the virtual and real photons was required to
be within the limits 5~mrad $<$
$\theta_{\gamma^{*}\gamma}$ $<$ 45~mrad. 

An event sample was selected requiring
that the squared missing-mass $M_{\textrm{X}}^{2}=(q+M_{p}-q')^{2}$
of the $e\,p \rightarrow\, e'\,\gamma\, \textrm{X}$ measurement
corresponded to the square of the proton mass, $M_{p}$, within the limits of the
energy resolution of the H{\sc ermes} spectrometer (mainly the calorimeter). Recall that $q$ is the
four-momentum of the virtual photon, $p$ is the initial four-momentum
of the target proton and $q'$ is the four-momentum of the produced
photon. The ``exclusive region'' was defined as $-$($1.5$\,GeV)$^{2} <
M_{\textrm{X}}^{2}$ $<$ (1.7\,GeV)$^{2}$, as in
ref.~\cite{Air09}. This exclusive region was shifted by up to
0.17\,GeV$^{2}$ for certain subsets of the data in order to reflect observed differences in the distributions of the electron and positron data samples~\cite{Bur10}. 
The data sample in the exclusive region contains events not only involving the production of real photons in which the proton remains intact, but also
events involving the excitation of the target proton to a $\Delta^+$
resonant state (``associated production''). The recoiling proton is not considered and the calorimeter resolution does not allow separation of all of the latter events from the rest of the data sample.
No systematic uncertainty is assigned for the contributions from these
events; they are treated as part of the signal. A Monte Carlo
calculation based on the parameterisation from ref.~\cite{Bra76} is
used to estimate the fractional contribution to the event sample from resonant
production in each kinematic bin; the uncertainty on this estimate
cannot be adequately quantified because no measurements have been made
in the H{\sc ermes} kinematic region. The results of the estimate, called the associated fractions and labelled ``Assoc. fraction'', are shown in the last row of figures~\ref{bsa_xbjrange}--\ref{bca_xbjrange2} in the results section. The method used to perform this estimation is described in detail in
ref.~\cite{Air08}.

\section{Experimental extraction of asymmetry amplitudes}

The expectation value of the experimental yield $N$ is parameterised as
\begin{equation}
 \langle N(e_{\ell},P_{\ell},\phi)\rangle =
\mathcal{L}(e_{\ell},P_{\ell})\eta(e_{\ell},\phi)\sigma_{\textrm{UU}}
(\phi)
[1+P_{\ell}\mathcal{A}_{\textrm{LU}}^{\textrm{DVCS}}(\phi)+e_{\ell}P_{\ell}
\mathcal{A}_{\textrm{LU}}^{\textrm{I}}(\phi)+e_{\ell}\mathcal{A}_{\textrm{C}}
(\phi)],
\end{equation}
where $P_\ell$ is the beam polarisation, $\mathcal{L}$ is the integrated luminosity, $\eta$ is the detection
efficiency and d$\sigma_{\textrm{UU}}$ denotes the
cross section for an unpolarised target summed over both beam charges and
beam helicities. The asymmetries $\mathcal{A}_{\textrm{LU}}^{\textrm{DVCS}}(\phi)$, $\mathcal{A}_{\textrm{LU}}^{\textrm{I}}(\phi)$ and
$\mathcal{A}_{\textrm{C}}(\phi)$ are expanded in
$\phi$ as
\begin{equation}
 \mathcal{A}_{\textrm{LU}}^{\textrm{DVCS}}(\phi) \simeq 
A_{\textrm{LU,DVCS}}^{\sin\phi}\sin\phi 
+ \sum^{1}_{n=0} A_{\textrm{LU,DVCS}}^{\cos(n\phi)}\cos(n\phi), 
\label{aludvcs_asym}
\end{equation}
\begin{equation}
\mathcal{A}_{\textrm{LU}}^{\textrm{I}}(\phi) \simeq \sum^{2}_{n=1}
A_{\textrm{LU,I}}^{\sin(n\phi)}\sin(n\phi) 
+ \sum^{1}_{n=0} A_{\textrm{LU,I}}^{\cos(n\phi)}\cos(n\phi), 
\label{alui_asym}
\end{equation}
\begin{equation}
\mathcal{A}_{\textrm{C}}(\phi) \simeq \sum^{3}_{n=0}
A_{\textrm{C}}^{\cos(n\phi)}\cos(n\phi) 
+ A_{\textrm{C}}^{\sin\phi}\sin\phi,
\label{ac_asym}
\end{equation}
where the approximation is due to the truncation of the infinite
Fourier series that would describe exactly the fitted distribution. Only the $\sin(n\phi)$ terms of the
$\mathcal{A}_{\textrm{LU}}$ asymmetries and the $\cos(n\phi)$ terms of the
$\mathcal{A}_{\textrm{C}}$ asymmetry are motivated by the
  physical processes under investigation. The other terms
are included both as a consistency check for any off-phase
extraneous harmonics in the data and as a test of the
normalisation of the fit. These terms are expected to be
consistent with zero and are found to be so.

A maximum-likelihood fitting technique \cite{Bar90} was used to
extract the asymmetry amplitudes in each kinematic bin of $-t$, $x_{\textrm{B}}$ and $Q^{2}$.
This method, described in ref.~\cite{Air08}, fits the expected
azimuthal distribution function to the data without introducing binning effects in $\phi$.
Event weights are introduced in the fitting procedure to account for
luminosity imbalances with respect to the beam charge and polarisation.

The asymmetry amplitudes $A_{\textrm{LU,I/DVCS}}^{\sin(n\phi)}$ and
$A_{\textrm{C}}^{\cos(n\phi)}$ relate respectively to the Fourier
coefficients $s_{\textrm{unp},n}^{W}$ and $c_{\textrm{unp},n}^{\textrm{I}}$ from the interference and DVCS terms in eqs.~\ref{e:alui}-\ref{e:ac}. The asymmetry amplitudes
may also be affected by the lepton propagators and the other
$\phi$-dependent terms in the denominators in
eqs.~\ref{e:alui}-\ref{e:ac}.

The DVCS asymmetry amplitude $A^{\sin\phi}_{\textrm{LU,DVCS}}$ receives a
contribution from the $\mathcal{C}_{\textrm{unp}}^{\textrm{DVCS}}$-function,
which is bilinear in CFFs. However, this twist-3 amplitude is
inherently small in H{\sc ermes} kinematic conditions due to the size of the $s_{\textrm{unp},1}^{\textrm{DVCS}}$ Fourier coefficient compared to the contributions from the $c_{\textrm{unp},n}^{\textrm{BH}}$ coefficients in the denominator of eq.~\ref{e:aludvcs}. As a result of the more complicated dependence on the CFFs and this suppression, it
is more difficult to constrain GPDs via the measurement of $A^{\sin\phi}_{\textrm{LU,DVCS}}$ than from the kinematically-unsuppressed leading twist amplitudes.

The leading-twist asymmetry amplitudes are $A_{\textrm{C}}^{\cos(0\phi)}$, $A_{\textrm{C}}^{\cos\phi}$ and $A_{\textrm{LU,I}}^{\sin\phi}$, which are proportional to the Fourier coefficients $c_{\textrm{unp},0}^{\textrm{I}}$, $c_{\textrm{unp},1}^{\textrm{I}}$ and $s_{\textrm{unp},1}^{\textrm{I}}$ defined in eqs.~\ref{eq:c0}--\ref{eq:s1}. Whilst all of these amplitudes receive contributions from $\mathcal{C}_{\textrm{unp}}^{\textrm{I}}$, $c_{\textrm{unp},0}^{\textrm{I}}$ is kinematically suppressed in comparison to $c_{\textrm{unp},1}^{\textrm{I}}$, so $A_{\textrm{C}}^{\cos\phi}$ and $A_{\textrm{LU,I}}^{\sin\phi}$ are expected to have the largest magnitude in H{\sc ermes} kinematic conditions.

Although strictly dependent on higher-twist quantities, the asymmetry amplitudes $A_{\textrm{LU},\textrm{I}}^{\sin(2\phi)}$ and $A^{\cos(2\phi)}_{\textrm{C}}$ can also be expressed as having a dependence on $\mathcal{C}_{\textrm{unp}}^{\textrm{I}}$ using the Wandzura-Wilzcek approximation~\cite{Wan}, i.e. neglecting antiquark-gluon-quark contributions; these amplitudes that are dependent on higher-twist objects can therefore be considered as being functionally similar, but kinematically suppressed, when compared to the amplitudes that are dependent only on leading-twist objects.

The $A^{\cos(3\phi)}_{\textrm{C}}$ amplitude depends on the
$c_{\textrm{unp},3}^{\textrm{I}}$ Fourier coefficient and hence the
$\mathcal{C}_{\textrm{T,unp}}^{\textrm{I}}$-function. Although the CFFs
in this function are of leading twist, they relate to gluon helicity-flip GPDs and are thus suppressed by $\alpha_{\textrm{S}}/\pi$, where $\alpha_{\textrm{S}}$ is the strong coupling constant. Table~\ref{tab_amplitudes} presents the asymmetry amplitudes extracted in this analysis and, for each of them, the
related dominant Fourier coefficient and $\mathcal{C}$-function, and the twist-level at which the contributing GPDs enter.
\begin{table}
\resizebox{\textwidth}{!} {
\begin{tabular}{|c|c|c|c|c|}
\hline
Asymmetry Amplitude& Fourier Coefficient& Dominant CFF Dependence & Twist-Level   \\ 
\hline
\hline
$A_{\textrm{LU,I}}^{\sin\phi}$ & $s_{\textrm{unp},1}^{\textrm{I}}$  &
$\mathfrak{Im}\mathcal{C}_{\textrm{unp}}^{\textrm{I}}$
&  2 \\
\hline
$A_{\textrm{LU,I}}^{\sin(2\phi)}$ & $s_{\textrm{unp},2}^{\textrm{I}}$ 
&
$\mathfrak{Im}\mathcal{C}_{\textrm{unp}}^{\textrm{I}}$
&  3 \\
\hline
\hline
$A_{\textrm{LU,DVCS}}^{\sin\phi}$ & $s_{\textrm{unp},1}^{\textrm{DVCS}}$ &
$\mathfrak{Im}\mathcal{C}_{\textrm{unp}}^{\textrm{DVCS}}$ &  3 \\
\hline
\hline
$A_{\textrm{C}}^{\cos(0\phi)}$ & $c_{\textrm{unp},0}^{\textrm{I}}$  &
$\mathfrak{Re}\mathcal{C}_{\textrm{unp}}^{\textrm{I}}$ & 2
\\
\hline
$A_{\textrm{C}}^{\cos\phi}$ & $c_{\textrm{unp},1}^{\textrm{I}}$  &
$\mathfrak{Re}\mathcal{C}_{\textrm{unp}}^{\textrm{I}}$ & 2
\\
\hline
$A_{\textrm{C}}^{\cos(2\phi)}$ & $c_{\textrm{unp},2}^{\textrm{I}}$ &
$\mathfrak{Re}\mathcal{C}_{\textrm{unp}}^{\textrm{I}}$ & 3 \\
\hline
$A_{\textrm{C}}^{\cos(3\phi)}$ & $c_{\textrm{unp},3}^{\textrm{I}}$ &
$\mathfrak{Re}\mathcal{C}_{\textrm{T,unp}}^{\textrm{I}}$ &  2 \\
\hline
 \end{tabular}
}
\caption{Asymmetry amplitudes that can
be extracted from the available data set, the related Fourier
coefficients, dominant $\mathcal{C}$-functions and twist-levels.}
\label{tab_amplitudes}
\end{table}
 
\section{Background corrections and systematic uncertainties for the
  2006-2007 data}
  
The extracted asymmetry amplitudes are subject to systematic uncertainties that result from a combination of background processes, shifts in the missing-mass distributions, and various detector and binning effects determined in the same manner as used in refs.~\cite{Air08,Air09} in order to maintain consistency with the results published in ref.~\cite{Air09} and therefore facilitate the combination of the two data sets. No systematic
uncertainty is assigned from the intermittent fault in the calorimeter
mentioned in section 2; the number of events in which the fault
occurred is very small for the 2006-2007 data sample and completely negligible in the context
of the combined data sets.

The contribution to the uncertainties on the amplitude measurements
arising from background in the data from neutral meson
production is predominantly due to the failure to identify one of
  the two photons from the decay of these neutral mesons. It is
    possible that both photons from the decay of a neutral meson could
    be boosted into a single calorimeter cluster and thus be
    reconstructed as a single photon produced in the BH or DVCS
    processes. It is similarly possible that the trajectory of one of the produced photons goes outwith the spectrometer acceptance and that the remaining photon is mistaken for one produced by the BH or DVCS processes. The $\mathcal{A}^{\textrm{DVCS}}_{\textrm{LU}}$ amplitudes are corrected for
  the fraction of the data sample and the magnitude of the asymmetry
  due to semi-inclusive pion production. A further uncertainty is assigned to
  these amplitudes for the influence of photons produced in decays of
  exclusively produced pions. Corrections for dilution of the amplitudes for the
    $\mathcal{A}_{\textrm{LU}}^{\textrm{I}}$ and $\mathcal{A}_{\textrm{C}}$
    asymmetries are applied. No asymmetry value is assigned to the
    influence of the dilution because influences from meson production are expected to
    vanish when considering the difference between beam charges.
The procedure for estimating the
uncertainty and correction factor for each measured amplitude value is described in detail in
refs.~\cite{Air08,Air09}. Each measurement in the $-t$, $x_{\textrm{B}}$ and
  $Q^{2}$ projections has this uncertainty estimated at the
  centre of the relevant kinematic bin and included as part of the
  total systematic uncertainty. 

A contribution to the systematic uncertainties of the
 measured amplitudes arises from shifts in  the
missing-mass distributions. Such shifts appear in a comparison of electron and positron data~\cite{Zei09,Bur10}. One quarter of the difference between the asymmetries extracted using the standard
and shifted missing-mass windows is taken as the corresponding systematic
uncertainty.

The predominant contribution to the systematic uncertainty arises from detector
effects. These include the acceptance of the spectrometer, smearing
effects due to detector resolution (e.g. the minimum opening angle
between the scattered lepton and produced photon trajectories that can
be resolved in the calorimeter), external radiation in detector
material, potential misalignment and the finite bin width of the $-t$,
$x_{\textrm{B}}$ and $Q^{2}$ projections. In order to quantify these effects, events were generated using a Monte Carlo simulation of the
spectrometer that included them. An event generator based on the GPD
model described in ref.~\cite{Guz06} was used for the simulation
because its output describes the data well and it was employed in
ref.~\cite{Air09}. Asymmetry amplitudes were extracted from these
simulated events using the same analysis procedure used to extract
amplitudes from experimental data. In each kinematic bin, the
systematic uncertainty was determined 
as the difference between the asymmetry amplitude reconstructed from the simulated data and that
calculated from the GPD model at the average $-t$, $x_{\textrm{B}}$ and
$Q^{2}$ value for that bin. The MC simulation shows that, in terms of kinematic
  smearing, the data sample is
  $99.9\%$  pure in each of the large ``overall'' bins. In the
  kinematic projections, the best purity is found in the sixth $Q^2$
  bin, which is $98\%$ pure. The least pure bin is the third bin in
  $-t$, where approximately one-third of the events reconstructed in this
  bin are generated from outwith it. The average 
  kinematic values in each of the bins are shown to be shifted by no more than $5\%$ as a result of
  kinematic smearing and the typical effect is a shift in the average
  kinematic values of a bin on the order of $1\%$.

The total systematic uncertainty for the 2006-2007 data sample was
determined for each kinematic bin by adding in quadrature the
  uncertainties arising from the background correction, the
  missing-mass shifts, and the detector effects.  The 1996-2005 sample also has a systematic uncertainty from misalignment of the spectrometer~\cite{Air09}, which has been eliminated for the 2006-2007 data sample due to improved
  surveying measurements. However, because the systematic uncertainty calculation for the 2006-2007 data uses the same Monte Carlo generator and reconstruction technique as was used for the 1996-2005 data, the systematic uncertainty for 2006-2007 is overestimated; this overestimate is very slight, because the systematic uncertainty contribution from potential misalignment affecting the asymmetries extracted from the 1996-2005 data set is very small~\cite{Bur10}.

Table~\ref{table_systematic_contributions_0607} shows, for each
physically-motivated amplitude extracted from the 2006-2007 data (in a
single, overall kinematic bin), the contribution of the various
systematic uncertainties. The beam polarisation measurements have total uncertainties of 2.8\% for the 1996-2005 data-taking period and 3.4\% for the 2006-2007 data taking period. These uncertainties are
present in the beam-helicity amplitudes and are, as independent
  scale uncertainties, not included in the other presented uncertainties.
  \begin{table}
\resizebox{\textwidth}{!} {
 \begin{tabular}{|c|c||c|c|c|c|c|}
  \hline
 & $A$ $\pm$ $\delta_{stat.}$& Background & Missing-Mass Shift  & Detector Effects & & Total $\delta_{syst.}$ \\
  \hline
  \hline
  $A_{\textrm{LU,I}}^{\sin\phi}$ & -0.222  $\pm$  0.023  & 0.002 & 0.001 & 0.022 & & 0.022 \\
  \hline
  $A_{\textrm{LU,DVCS}}^{\sin\phi}$ & 0.005  $\pm$  0.023  & 0.002 & 0.002 & 0.001 & & 0.003 \\
  \hline
  $A_{\textrm{LU,I}}^{\sin(2\phi)}$ & 0.005  $\pm$  0.023  & 0.003 & 0.001 & 0.001 & & 0.003 \\
  \hline
  \hline
  $A_{\textrm{C}}^{\cos(0\phi)}$ & -0.024 $\pm$  0.004 & 0.001 & 0.003 & 0.010 & & 0.011 \\
  \hline
  $A_{\textrm{C}}^{\cos\phi}$ & 0.032  $\pm$  0.006 & 0.002 & 0.001 & 0.001 & & 0.002 \\
  \hline
  $A_{\textrm{C}}^{\cos(2\phi)}$ & -0.004  $\pm$  0.005 & 0.001 & 0.000 & 0.014 & & 0.014 \\
  \hline
  $A_{\textrm{C}}^{\cos(3\phi)}$ & 0.001  $\pm$   0.005 & 0.000 & 0.001 & 0.003 & & 0.004 \\
  \hline
 \end{tabular}
}
  \caption{The values of the physically-motivated asymmetries extracted in a single bin over all kinematic variables with their statistical uncertainties are presented in the second column of this table. The third, fourth and fifth columns show the contributions to the overall systematic uncertainties of the extracted asymmetry amplitudes due to the background correction, the time-dependent shifts of the missing-mass distributions and detector effects for the 2006-2007 data. The total
systematic uncertainties of the amplitudes, shown in the right-most column, are the individual contributions added in quadrature.}
  \label{table_systematic_contributions_0607}
\end{table}

\section{Results}

\begin{figure}
\centering
\includegraphics[width=\textwidth,keepaspectratio]{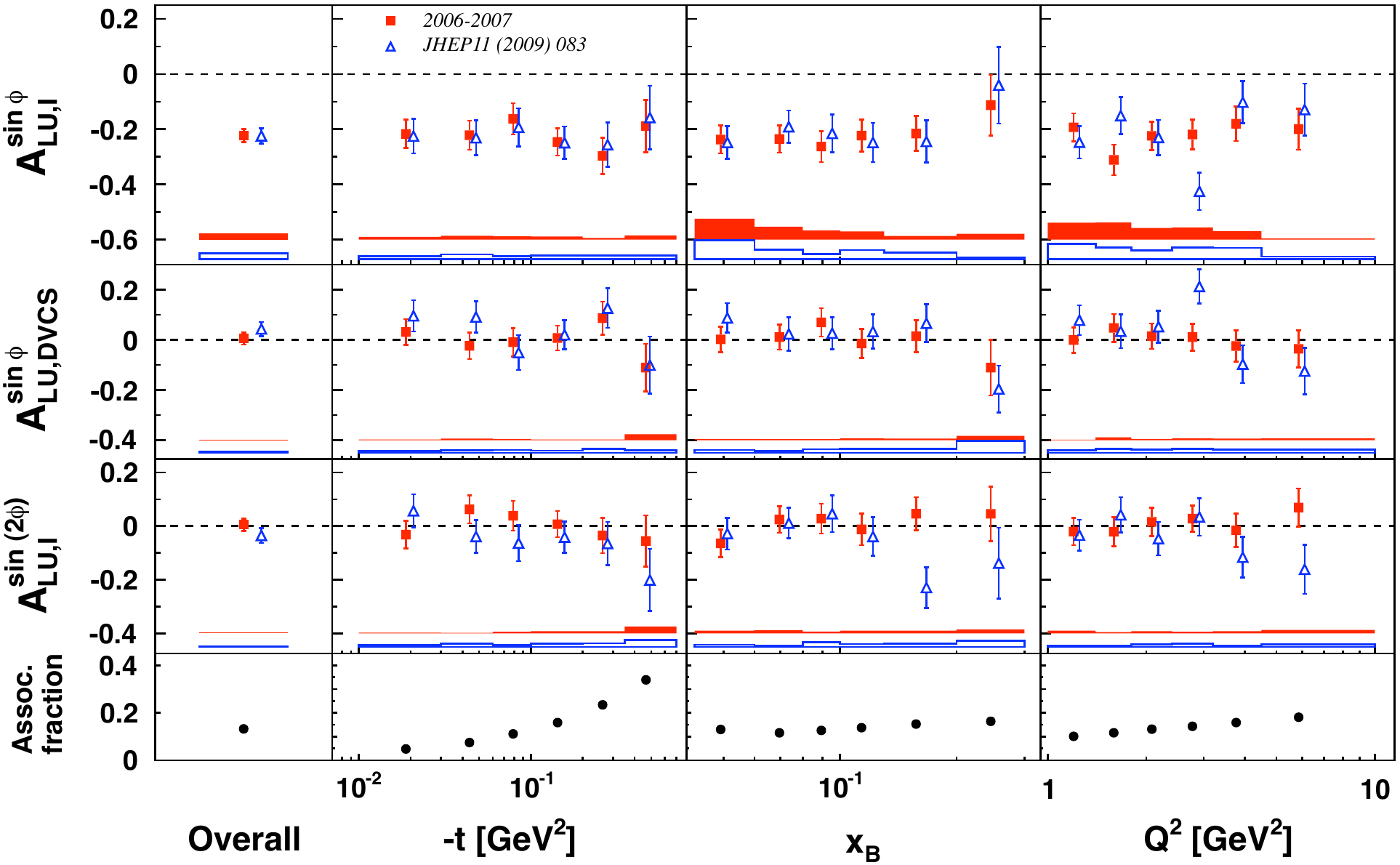}
  \caption{Beam-helicity asymmetry amplitudes extracted separately from
the unpolarised 1996-2005 (open triangles)~\cite{Air09} and 2006-2007 (filled squares)
hydrogen data. The error bars represent the statistical uncertainties. The error bands represent the systematic uncertainties.
An additional 2.8\,\% and 3.4\,\% scale uncertainty for the 1996-2005 and
2006-2007 data respectively is present in the amplitudes due to the uncertainty of
the beam polarisation measurement. The simulated fractional contribution from associated production to the yield in each kinematic bin is shown in the bottom row.}
 \label{release_bsa_0607}
\end{figure}

\begin{figure}
\centering
 \includegraphics[width=\textwidth,keepaspectratio]{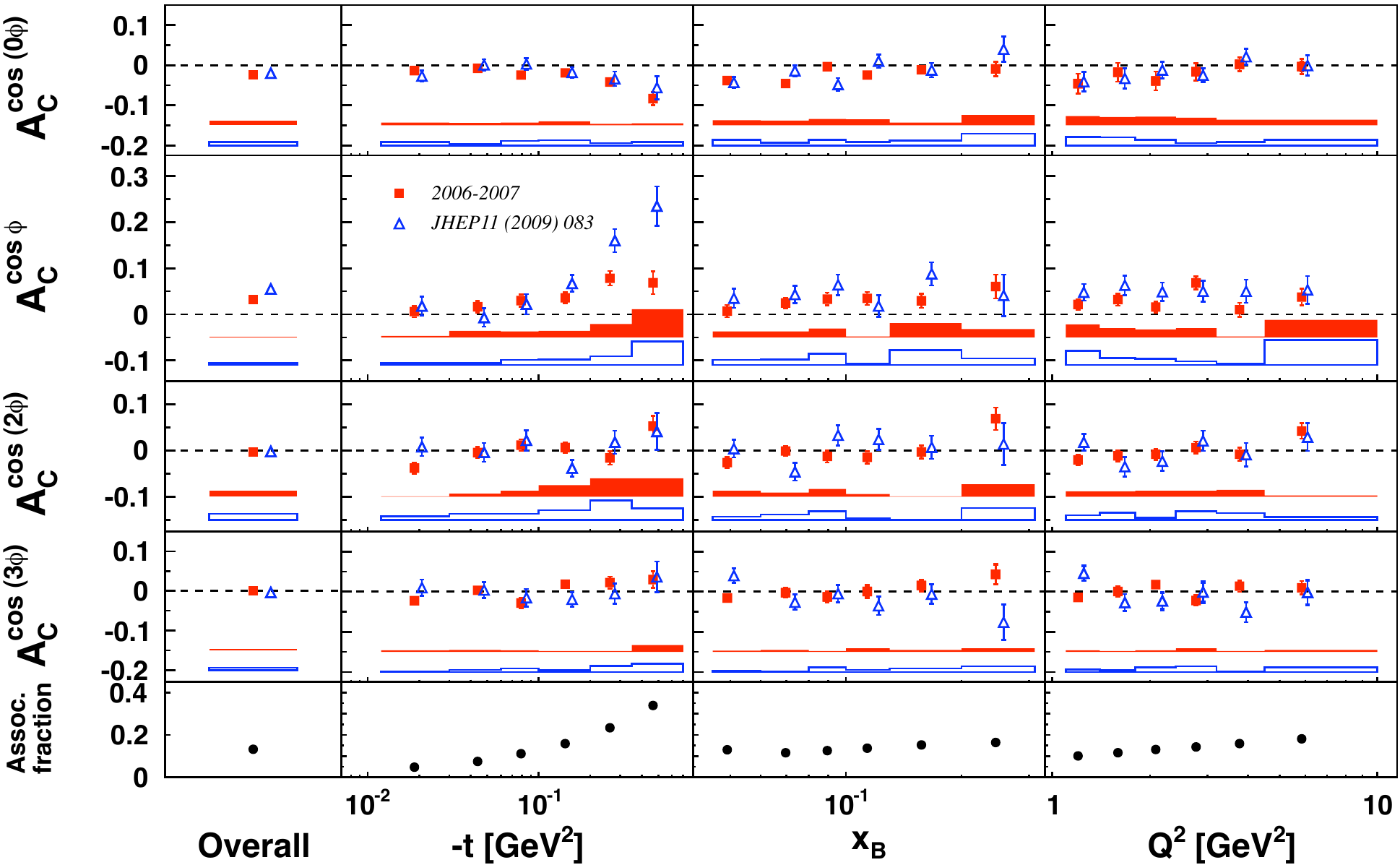}
  \caption{Beam-charge asymmetry amplitudes extracted separately from the unpolarised 1996-2005 (open triangles)~\cite{Air09} and 2006-2007 (filled squares) hydrogen data.
The error bars represent the statistical uncertainties. The error bands represent the systematic uncertainties. The simulated fractional contribution from associated production to the yield in each kinematic bin is shown in the bottom row.}
 \label{release_bca_0607} 
 \end{figure}

In figures~\ref{release_bsa_0607} and~\ref{release_bca_0607}, results for the beam-helicity and beam-charge asymmetry amplitudes extracted from the 2006-2007 hydrogen data sets in this work are compared with results extracted from the 1996-2005\footnote{The hydrogen gas target for the data set of 1996-2005 was either unpolarised, transversely polarised or longitudinally polarised. However, the time-averaged polarisation of the polarised targets was negligible, while the rapid reversal time (60--180s) of the polarisation direction minimised bias due to detector effects.} data set published previously~\cite{Air09}. Each of the asymmetry amplitudes is shown extracted in one bin over all kinematic variables (``Overall'') and also projected against $-t$, $x_{\textrm{B}}$ and $Q^{2}$. The beam-helicity asymmetry amplitudes are subject to an additional scale uncertainty from the measurement of the beam polarisation, which is stated in the captions of the figures.

A statistical test (Student's t-test) was applied in order to check for possible incompatibility between the asymmetry amplitudes extracted from the two data sets. Only the statistical uncertainties were employed in this test as the largest contributions to the systematic uncertainties, i.e. effects from detector resolution, acceptance, misalignment and smearing, are largely correlated. This test revealed no significant evidence for incompatibility between the data sets. The beam-helicity and beam-charge asymmetry amplitudes can therefore be extracted from the entire hydrogen data set recorded during the entire experimental operation of H{\sc ermes}.

\begin{figure}
 \centering
 \includegraphics[width=\textwidth]{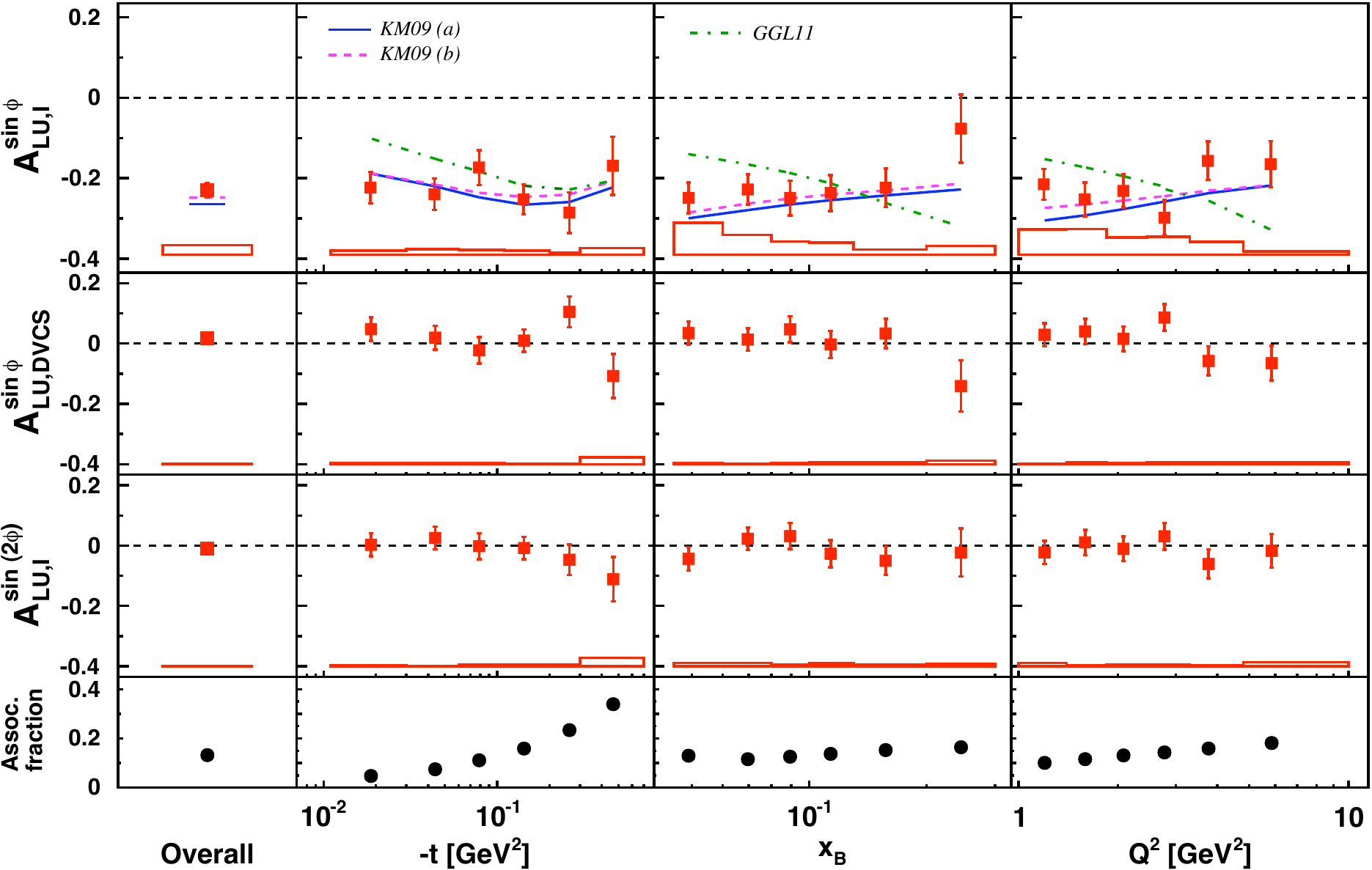}
  \caption{The $A_{\textrm{LU,I}}^{\sin\phi}$, $A_{\textrm{LU,DVCS}}^{\sin\phi}$ and
$A_{\textrm{LU,I}}^{\sin(2\phi)}$ beam-helicity asymmetry amplitudes extracted from all the unpolarised hydrogen data recorded at H{\sc ermes}
from 1996 until 2007. The error bars (bands) represent the statistical
(systematic) uncertainties. An additional 3.2\,\% scale uncertainty is present in the amplitudes due to the uncertainty of
the beam polarisation measurement. Solid and dashed lines (KM09) show model calculations from ref.~\cite{Kum09}; calculations from ref.~\cite{Liu11} are shown as dashed-dotted lines (GGL11). See text for details. The simulated fractional contribution from associated production to the yield in each kinematic bin is shown in the bottom row.}
  \label{bsa_xbjrange}
\end{figure}

\begin{figure}
\centering
    \includegraphics[width=\textwidth]{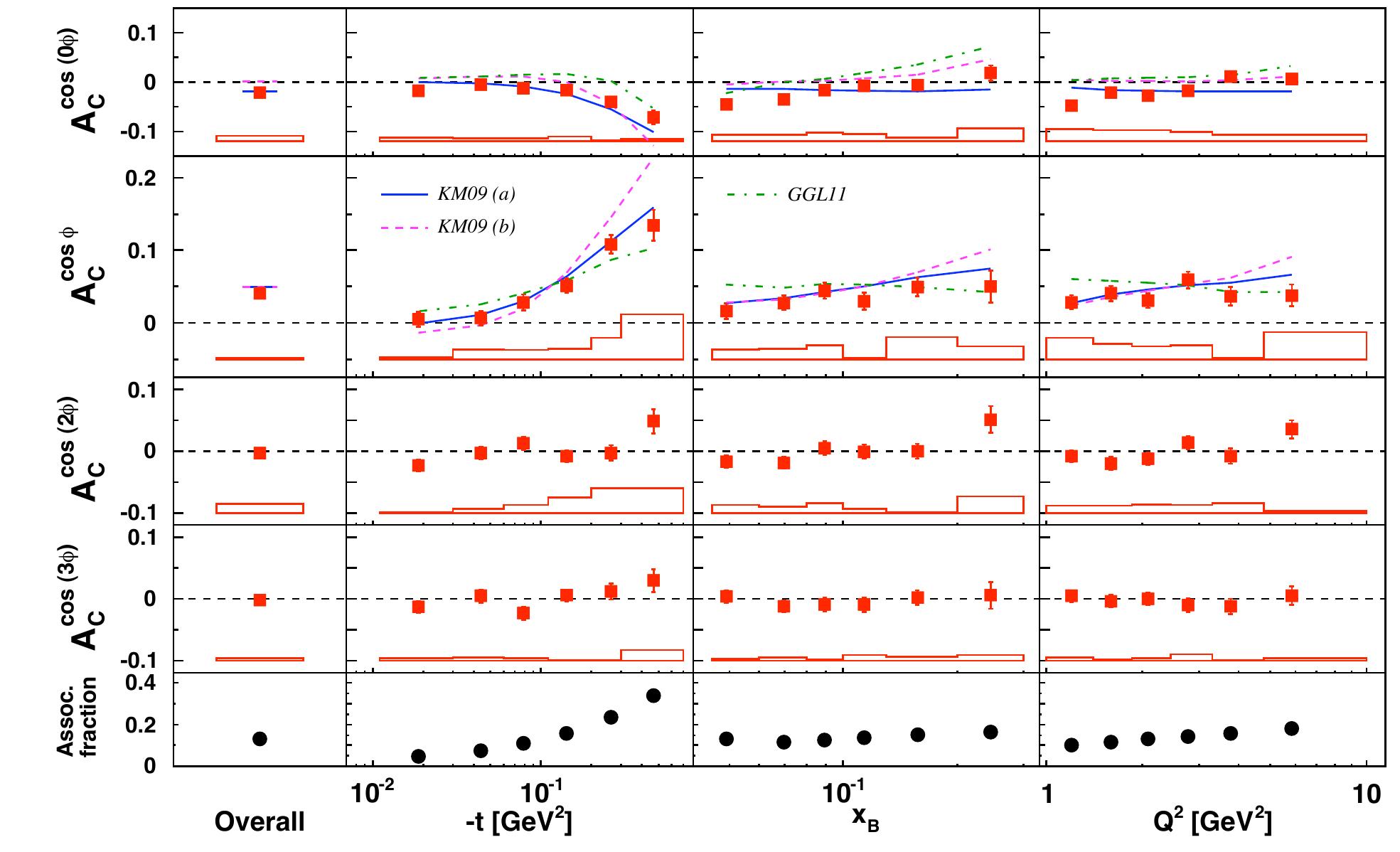}
    \caption{The $A_{\textrm{C}}^{\cos(0\phi)}$, $A_{\textrm{C}}^{\cos\phi}$, $A_{\textrm{C}}^{\cos(2\phi)}$ and $A_{\textrm{C}}^{\cos(3\phi)}$ beam-charge asymmetry amplitudes extracted from all the unpolarised hydrogen data recorded at H{\sc ermes} from 1996 until 2007. The error bars (bands) represent the statistical (systematic) uncertainties.  Theoretical calculations from the model described in ref.~\cite{Kum09} are shown as solid and dashed lines (KM09); calculations from ref.~\cite{Liu11} are shown as dashed-dotted lines (GGL11). See text for details. The simulated fractional contribution from associated production to the yield in each kinematic bin is shown in the bottom row.}
  \label{bca_xbjrange}
\end{figure}

\begin{figure}
\centering
 \includegraphics[width=\textwidth]{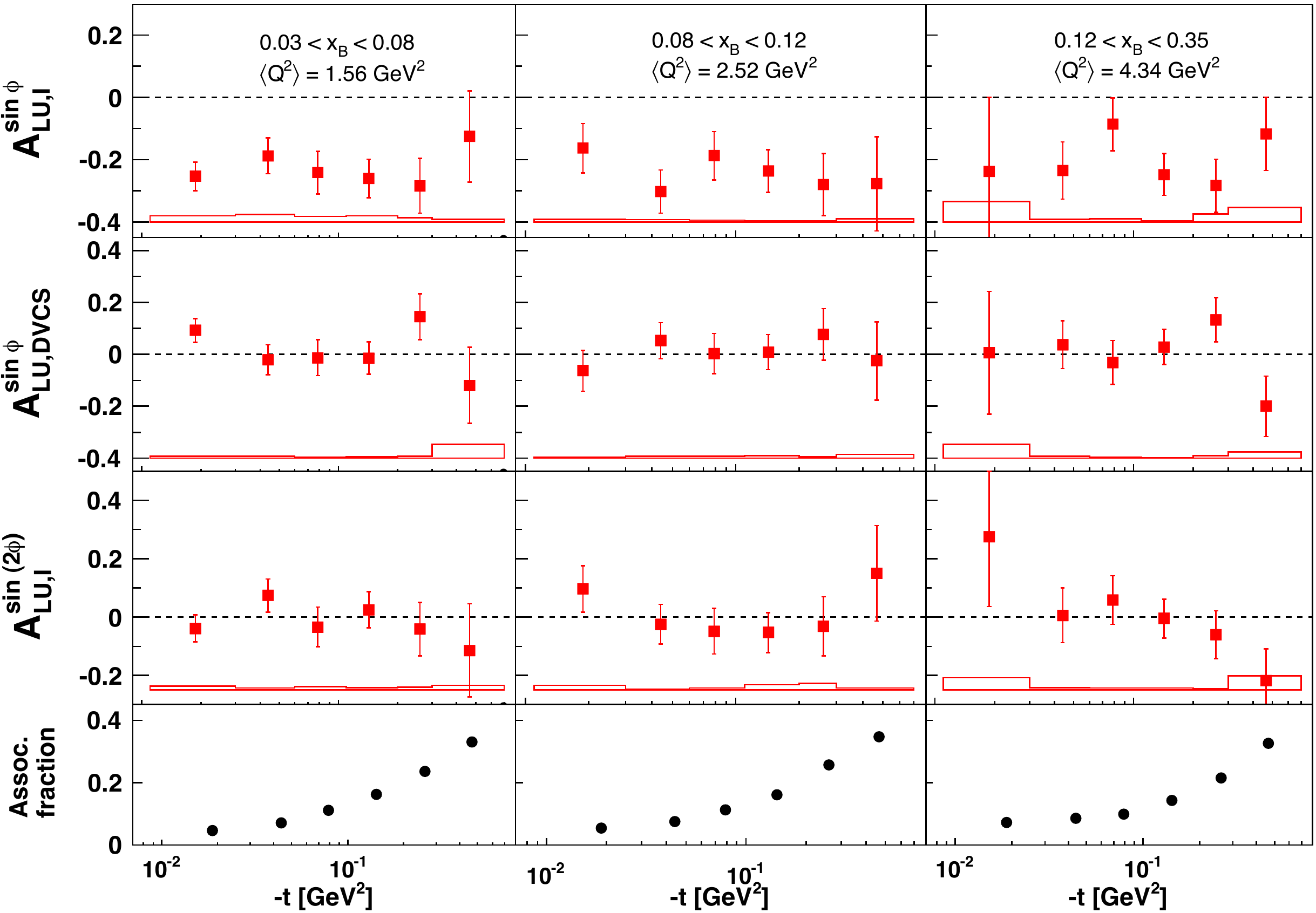}
  \caption{The $A_{\textrm{LU,I}}^{\sin\phi}$, $A_{\textrm{LU,DVCS}}^{\sin\phi}$ and
$A_{\textrm{LU,I}}^{\sin(2\phi)}$ beam-helicity asymmetry amplitudes
extracted from all the unpolarised hydrogen data recorded at H{\sc ermes} from 1996 until 2007 as a function of $-t$ for three different $x_{\textrm{B}}$ ranges. The error bars (bands) represent the statistical (systematic) uncertainties. An additional 3.2\,\% scale uncertainty is present in the amplitudes due to the uncertainty of the beam polarisation measurement. The simulated fractional contribution from associated production to the yield in each kinematic bin is shown in the bottom row.}
  \label{bsa_xbjrange2}
\end{figure}

\begin{figure}
\centering
    \includegraphics[width=\textwidth]{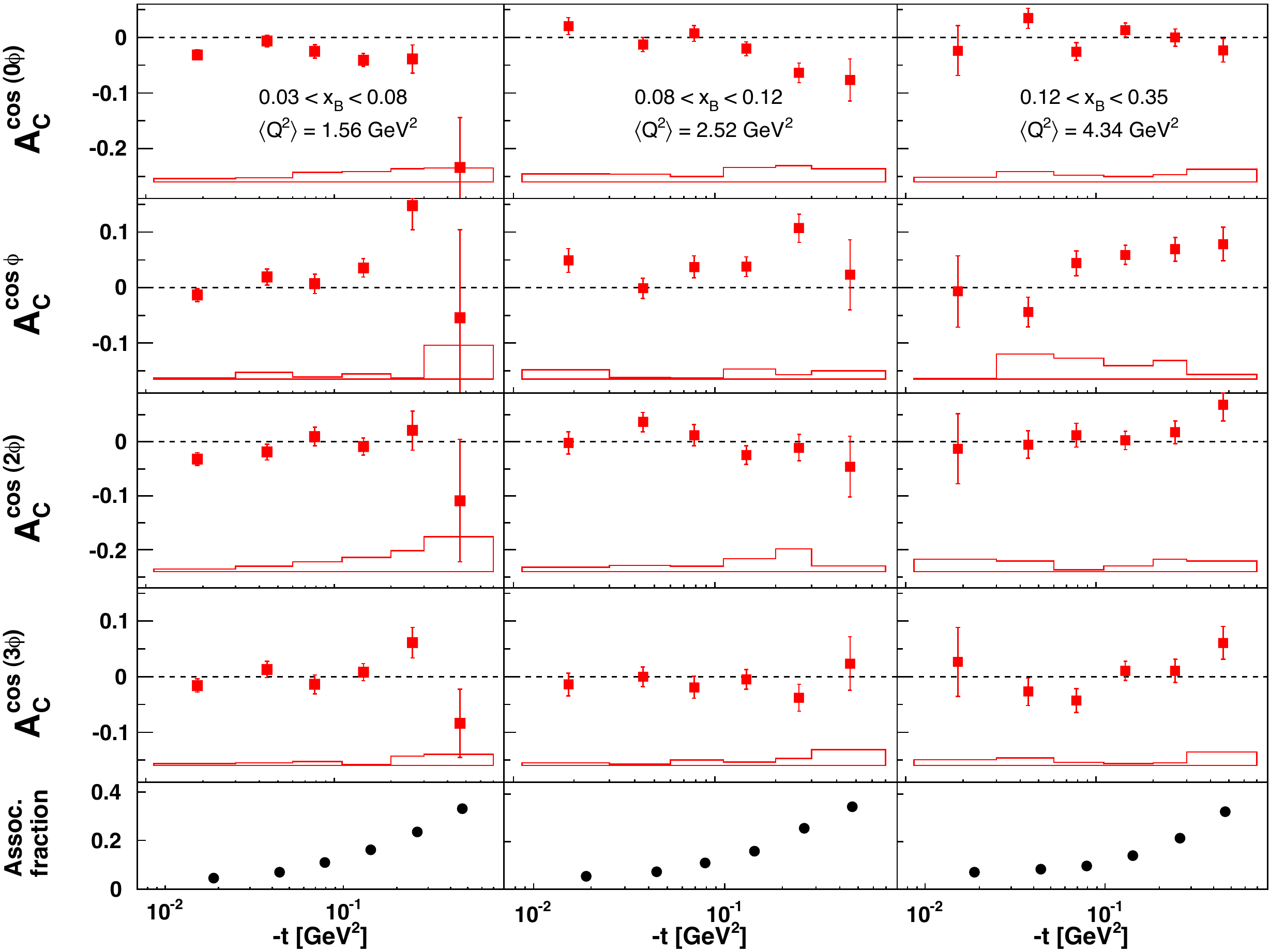}
    \caption{The $A_{\textrm{C}}^{\cos(0\phi)}$, $A_{\textrm{C}}^{\cos\phi}$, $A_{\textrm{C}}^{\cos(2\phi)}$ and $A_{\textrm{C}}^{\cos(3\phi)}$ beam-charge asymmetry amplitudes extracted from all the unpolarised hydrogen data recorded at H{\sc ermes} from 1996 until 2007 as a function of $-t$ for three different $x_{\textrm{B}}$ ranges. The error bars (bands) represent the statistical (systematic) uncertainties. The simulated fractional contribution from associated production to the yield in each kinematic bin is shown in the bottom row.}
  \label{bca_xbjrange2}
\end{figure}

The results of the beam-helicity and beam-charge asymmetry amplitudes extracted from the complete 1996-2007 hydrogen sample are shown in figures~\ref{bsa_xbjrange} and \ref{bca_xbjrange}. The number of analysable events available from the 2006-2007 data set (70352) is approximately three times greater than the number of events recorded in the 1996-2005 sample (24817). The  asymmetry amplitudes extracted from the complete 1996-2007 data set thus resemble the 2006-2007 result. This resemblance is not so evident for the beam-helicity asymmetry amplitudes as it is for the beam-charge asymmetry amplitudes, because the beam polarisation was lower in 2006 and 2007 than in 1996-2005 and thus the 2006-2007 data has a lower weighting in the combined fit. The major contributions to the systematic error bands associated with the asymmetry amplitudes extracted from the combined data set were determined using Monte Carlo simulations as explained in section~4, i.e. contributions from acceptance, smearing, finite bin widths and misalignment. The background in the combined data sample is estimated using the method from refs.~\cite{Air08,Zei09, Bur10}. The uncertainty contributions due to the observed shifts in the missing-mass distributions for the combined data sets were calculated using the procedure described in section~4 and the results were averaged. The total systematic uncertainties for the combined results are obtained by adding these three independent contributions in quadrature.

The first and second harmonics of $\mathcal{A}^{\textrm{I}}_{\textrm{LU}}$, which are sensitive to the interference term in the scattering amplitude, are shown in the first and third rows of figure~\ref{bsa_xbjrange}. The leading-twist amplitude $A_{\textrm{LU,I}}^{\sin\phi}$ has the largest magnitude of any of the amplitudes when extracted in a single bin from the entire data set. This amplitude shows no strong dependence on $-t$, $x_{\textrm{B}}$ or $Q^{2}$, implying a strong dependence at smaller values of $-t$ as the amplitude must approach zero as $-t$ approaches its minimum value because of the dependence of the amplitude on the factor $K$ defined in eq.~\ref{eq:K}. The $A_{\textrm{LU,I}}^{\sin\phi}$ amplitude is sensitive to the imaginary part of the CFF $\mathcal{H}$ and thereby can constrain GPD $\textit{H}$. The $A_{\textrm{LU,DVCS}}^{\sin\phi}$ asymmetry is shown in the second row of figure~\ref{bsa_xbjrange}. Both the $A_{\textrm{LU,DVCS}}^{\sin\phi}$ asymmetry amplitude and the $A_{\textrm{LU,I}}^{\sin(2\phi)}$ asymmetry amplitude are compatible with zero, and neither asymmetry amplitude shows any dependence on $-t$, $x_{\textrm{B}}$ or $Q^{2}$.

The $A_{\textrm{C}}^{\cos(n\phi)}$ amplitudes are shown in figure~\ref{bca_xbjrange}. The systematic uncertainties are estimated using the same procedure as was used to estimate those for the beam-helicity asymmetries. The leading twist $A_{\textrm{C}}^{\cos(0\phi)}$ and $A_{\textrm{C}}^{\cos\phi}$ amplitudes are both non-zero. There is an expected relationship between $A_{\textrm{C}}^{\cos(0\phi)}$ and $A_{\textrm{C}}^{\cos\phi}$ as they depend on the same $\mathcal{C}$-function. The kinematic suppression of $A_{\textrm{C}}^{\cos(0\phi)}$ with regard to $A_{\textrm{C}}^{\cos\phi}$ is approximately fulfilled. The measured amplitudes are found to diverge with opposite sign from zero at increasing values of $-t$ and they indicate a weak increase with $x_{\textrm{B}}$ and $Q^{2}$. The $A_{\textrm{C}}^{\cos(2\phi)}$ and $A_{\textrm{C}}^{\cos(3\phi)}$ amplitudes are both consistent with zero over the whole range in $-t$, $x_{\textrm{B}}$ and $Q^{2}$. The $A_{\textrm{C}}^{\cos(2\phi)}$ amplitude is related to twist-3 GPDs and $A_{\textrm{C}}^{\cos(3\phi)}$ relates to gluon helicity-flip GPDs. Both of these amplitudes are expected to be suppressed at H{\sc ermes} kinematic conditions compared to the leading twist amplitudes. 

The curves in  figures~\ref{bsa_xbjrange} and~\ref{bca_xbjrange} show the results of model calculations at the average value of the kinematic bins used for the data analysis. The solid and dashed curves show results of calculations from a global fit of GPDs to experimental data~\cite{Kum09} including information from H{\sc ermes} and Jefferson Lab., and the collider experiments at H{\sc era}. The basic model~\cite{Mue05,Kum07,Kum08} is a flexible GPD representation that is based on both a Mellin-Barnes integral and dispersion integral representation with weakly entangled skewness and $t$ dependences. The solid curves represent the model fit without data from the Jefferson Lab Hall A Collaboration~\cite{Cam06}; the model fit represented by the dashed curves includes these data. Both fits include the 1996-2005 H{\sc ermes} data. The model incorporates only twist-2 GPDs and so can provide results only for the $A_{\textrm{LU,I}}^{\sin\phi}$, $A_{\textrm{C}}^{\cos(0\phi)}$ and $A_{\textrm{C}}^{\cos\phi}$ asymmetry amplitudes. All of the relevant amplitudes reported here are well described by the model.

The dash-dotted curves in figures~\ref{bsa_xbjrange} and~\ref{bca_xbjrange} show the result of calculations from a fit based on a quark-diquark model with a Regge-inspired term that is included in order to describe accurately parton distribution functions at low $x$ values~\cite{Liu11}. The ``Regge'' term is extended to include contributions that determine the $t$-dependence of the corresponding GPD. The model incorporates fits to global deep-inelastic and elastic scattering data (to account for the $\xi$-independent limits and moments of the underlying GPDs) and DVCS data from Jefferson Lab. (to describe the skewness dependence). It describes the $t$-projections of the $A^{\sin\phi}_{\textrm{LU}}$ amplitude reported here well, but the projections in the other kinematic variables are not as well described. The model describes the trends of the $A_{\textrm{C}}^{\cos(0\phi)}$ and $A_{\textrm{C}}^{\cos\phi}$ asymmetry amplitudes well.

In order to provide more detailed information that can be used in future fits, in particular for the determination of the entanglement of the skewness and $-t$ dependences of GPDs, the amplitudes already presented in figures~\ref{bsa_xbjrange} and~\ref{bca_xbjrange} are shown as a function of $-t$ for three different ranges of $x_{\textrm{B}}$ in figures~\ref{bsa_xbjrange2} and~\ref{bca_xbjrange2}. These figures represent the kinematic dependences of the amplitudes in a less-correlated manner than the one-dimensional projections: within experimental uncertainty, there is no evidence of a correlation between the $-t$ and $x_{\textrm{B}}$ dependences for any of the amplitudes. 

The results from this paper will be made available in the Durham Database. The results will also be made available in the same 4-bin format as used in previous analyses at H{\sc ermes}~\cite{Air08,Air09,Air10b}.

\section{Summary}

Beam-helicity and beam-charge asymmetries in the azimuthal
distribution of real photons from hard exclusive leptoproduction on an unpolarised
hydrogen target have been presented. These asymmetries were extracted
from an unpolarised hydrogen data set taken during the 2006 and 2007
operating periods of H{\sc ermes}. Analogous asymmetry amplitudes were
extracted previously from hydrogen data obtained during the 1996-2005
experimental period as described in ref.~\cite{Air09}. A comparison of
the amplitudes extracted from these independent data sets has shown
that they are compatible and the asymmetry amplitudes can therefore be
extracted from the complete 1996-2007 data set. The asymmetry
amplitudes extracted from the complete data set are the most
statistically precise DVCS measurements presented by H{\sc
  ermes}. There is a strong signal in the first harmonic of the
interference contribution to the beam-helicity asymmetry. There are
non-zero amplitudes in the zeroth and first harmonics of the
beam-charge asymmetry. All asymmetry amplitudes related to higher
harmonics are consistent with zero. The results from the complete data
set are compared to calculations from ongoing work to fit GPD models to experimental data. All asymmetry amplitudes are also presented as projections in $-t$ in bins of $x_{\textrm{B}}$. No additional features are observed in any particular $x_{\textrm{B}}$-bin.

\acknowledgments

We gratefully acknowledge the \desy\ management for its support and the staff
at \desy\ and the collaborating institutions for their significant effort.
This work was supported by 
the Ministry of Economy and the Ministry of Education and Science of Armenia;
the FWO-Flanders and IWT, Belgium;
the Natural Sciences and Engineering Research Council of Canada;
the National Natural Science Foundation of China;
the Alexander von Humboldt Stiftung,
the German Bundesministerium f\"ur Bildung und Forschung (BMBF), and
the Deutsche Forschungsgemeinschaft (DFG);
the Italian Istituto Nazionale di Fisica Nucleare (INFN);
the MEXT, JSPS, and G-COE of Japan;
the Dutch Foundation for Fundamenteel Onderzoek der Materie (FOM);
the Russian Academy of Science and the Russian Federal Agency for 
Science and Innovations;
the U.K.~Engineering and Physical Sciences Research Council, 
the Science and Technology Facilities Council,
and the Scottish Universities Physics Alliance;
the U.S.~Department of Energy (DOE) and the National Science Foundation (NSF);
the Basque Foundation for Science (IKERBASQUE) and the UPV/EHU under program UFI 11/55;
and the European Community Research Infrastructure Integrating Activity
under the FP7 "Study of strongly interacting matter (HadronPhysics2, Grant
Agreement number 227431)".

\newpage
\appendix
\setcounter{equation}{0}

\section{Tables of results}

\begin{table}[width=15cm]
\resizebox{16cm}{!} {
\begin{tabular}{|c|c|c|c|c|r|r|r|} \hline
\multicolumn{2}{|c|}{} & $\langle -t\rangle$ & $\langle
x_{\text{B}}\rangle$ & $\langle Q^2 \rangle $ & 
\multicolumn{1}{c|}{\multirow{2}{*}{$A_{\text{LU,I}}^{\sin \phi} \pm \delta_{stat.} \pm \delta_{syst.}$}} & 
\multicolumn{1}{c|}{\multirow{2}{*}{$A_{\text{LU,DVCS}}^{\sin \phi} \pm \delta_{stat.} \pm \delta_{syst.}$}} & 
\multicolumn{1}{c|}{\multirow{2}{*}{$A_{\text{LU,I}}^{\sin (2\phi)} \pm \delta_{stat.} \pm \delta_{syst.}$}} \\ 
\multicolumn{2}{|c|}{} &  $[\text{GeV}^2]$ & & $[\text{GeV}^2]$ & &  &  \\
\hline \hline
\multicolumn{2}{|c|}{overall} &  0.117 & 0.097 &  2.52 &  -0.222  $\pm$  0.023  $\pm$   0.022 &
 0.005  $\pm$  0.023  $\pm$  0.003 & 0.005  $\pm$  0.023  $\pm$   0.003 \\
\hline
\multirow{6}{*}{\rotatebox{90}{\mbox{$-t [\text{GeV}^2]$}}} & 0.00-0.03 &  0.018 & 0.068 &  1.72 &  -0.217  $\pm$  0.051  $\pm$   0.010 &
 0.031  $\pm$  0.051   $\pm$  0.003 & -0.032  $\pm$  0.051  $\pm$   0.003\\
& 0.03-0.06 &  0.043 & 0.088 &  2.26&  -0.222 $\pm$   0.052   $\pm$  0.014 &
 -0.024 $\pm$   0.052  $\pm$   0.006 & 0.062  $\pm$  0.052  $\pm$   0.002\\
& 0.06-0.10 &  0.078 & 0.099 &  2.51 & -0.163 $\pm$   0.057   $\pm$  0.012 &
 -0.010  $\pm$  0.056  $\pm$   0.005 & 0.039  $\pm$  0.056   $\pm$  0.006 \\
& 0.10-0.20 &  0.142 & 0.110 &  2.79 &  -0.246 $\pm$   0.049  $\pm$   0.011 &
0.007  $\pm$  0.049  $\pm$   0.003 & 0.007  $\pm$  0.049  $\pm$  0.008\\
& 0.20-0.35 &  0.260 & 0.121 &  3.27 &  -0.297 $\pm$   0.066  $\pm$   0.006 &
0.086  $\pm$  0.066  $\pm$   0.003 & -0.035 $\pm$   0.066   $\pm$  0.008\\
& 0.35-0.70 &  0.460 & 0.125 &  3.82 &  -0.189  $\pm$  0.095  $\pm$   0.015 & 
-0.111  $\pm$  0.095   $\pm$  0.024 & -0.056 $\pm$   0.096  $\pm$   0.029\\
\hline
\multirow{6}{*}{\rotatebox{90}{\mbox{$x_{\text{B}}$}}} & 0.03-0.06 &  0.095 & 0.049 &  1.34 &  -0.237  $\pm$  0.050  $\pm$   0.076 &
0.002 $\pm$   0.050  $\pm$   0.005 & -0.064  $\pm$  0.051  $\pm$   0.010\\
& 0.06-0.08 &  0.091 & 0.069 &  1.80 &  -0.235  $\pm$  0.050  $\pm$   0.047 &
0.010 $\pm$  0.050  $\pm$   0.004 & 0.024 $\pm$   0.049  $\pm$   0.012\\
& 0.08-0.10 &  0.104 & 0.089 &  2.30 &  -0.263 $\pm$  0.057  $\pm$   0.033 &
0.069 $\pm$   0.056  $\pm$   0.004 & 0.028  $\pm$  0.056  $\pm$   0.007\\
& 0.10-0.13 &  0.121 &  0.113 &  2.93 &  -0.223  $\pm$  0.059   $\pm$  0.030 & 
-0.015  $\pm$  0.058  $\pm$   0.007 & -0.012  $\pm$  0.059  $\pm$   0.010\\
& 0.13-0.20 &  0.159 & 0.157 &  4.06&  -0.216  $\pm$  0.063  $\pm$   0.013 &
0.014  $\pm$  0.063  $\pm$   0.006 & 0.046  $\pm$  0.061  $\pm$   0.010 \\
& 0.20-0.35 &  0.231 & 0.244 &  6.14 &  -0.113 $\pm$ 0.110  $\pm$   0.021 &
-0.111  $\pm$  0.110 $\pm$    0.017 & 0.046  $\pm$  0.102  $\pm$  0.015\\
\hline
\multirow{6}{*}{\rotatebox{90}{\mbox{$Q^2 [\text{GeV}^2]$}}} & 1.00-1.40 &  0.076 & 0.054  & 1.20 &  -0.193  $\pm$  0.051  $\pm$   0.061 &
-0.001 $\pm$   0.051  $\pm$   0.002 & -0.020  $\pm$  0.051   $\pm$  0.010 \\
& 1.40-1.80 &  0.089 & 0.069 &  1.59 &  -0.311 $\pm$  0.055  $\pm$   0.062 &
0.047  $\pm$  0.055  $\pm$   0.011 & -0.021 $\pm$   0.054  $\pm$   0.005\\
& 1.80-2.40 &  0.104 & 0.085 &  2.08 &  -0.224 $\pm$   0.051  $\pm$   0.042 &
0.014 $\pm$   0.051  $\pm$   0.005 & 0.015  $\pm$  0.053  $\pm$   0.008\\
& 2.40-3.20 &  0.126 & 0.105  & 2.77 &  -0.219 $\pm$   0.054  $\pm$   0.044 &
0.010  $\pm$  0.054 $\pm$    0.007 & 0.028   $\pm$ 0.049  $\pm$   0.006\\
& 3.20-4.50 &  0.151 & 0.134 &  3.76 &  -0.180 $\pm$   0.063  $\pm$   0.031 &
-0.025  $\pm$  0.062 $\pm$    0.006 & -0.016 $\pm$   0.062  $\pm$   0.008\\
& 4.50-10.0 &  0.218 & 0.200 &  5.82 &  -0.200  $\pm$  0.074 $\pm$    0.004 &
-0.036  $\pm$  0.074  $\pm$   0.007 & 0.069 $\pm$  0.071 $\pm$  0.013\\
\hline
  \end{tabular}
}
\caption{Results of the $A_{\textrm{LU,I}}^{\sin(n\phi)}$ and
  $A_{\textrm{LU,DVCS}}^{\sin \phi}$ asymmetry amplitudes with
  statistical and systematic uncertainties and the average values of the kinematic variables from unpolarised hydrogen target data taken during 2006-2007 at H{\sc ermes} for each $-t$, $x_{\textrm{B}}$ and $Q^{2}$ bin. An additional 3.4\,\% scale uncertainty is present in the amplitudes due to the uncertainty of the beam polarisation measurement.}
\end{table}

\begin{table}[width=15cm]
\resizebox{16cm}{!} {
\begin{tabular}{|c|c|c|c|c|r|r|r|r|} \hline
\multicolumn{2}{|c|}{} & $\langle -t\rangle$ & $\langle
x_{\text{B}}\rangle$ & $\langle Q^2 \rangle $ & 
\multicolumn{1}{c|}{\multirow{2}{*}{$A_{\text{C}}^{\cos (0\phi)}\pm \delta_{stat.} \pm \delta_{syst.}$ }} & 
\multicolumn{1}{c|}{\multirow{2}{*}{$A_{\text{C}}^{\cos \phi } \pm \delta_{stat.} \pm \delta_{syst.}$}} & 
\multicolumn{1}{c|}{\multirow{2}{*}{$A_{\text{C}}^{\cos (2\phi) } \pm \delta_{stat.} \pm \delta_{syst.}$ }} &
\multicolumn{1}{c|}{\multirow{2}{*}{$A_{\text{C}}^{\cos (3\phi) } \pm \delta_{stat.} \pm \delta_{syst.}$}} \\ 
\multicolumn{2}{|c|}{} &  $[\text{GeV}^2]$ & & $[\text{GeV}^2]$ & &  & &  \\
\hline
\hline
\multicolumn{2}{|c|}{overall} &  0.117 & 0.097 &  2.52 &  -0.024 $\pm$  0.004 $\pm$  0.011 & 
0.032  $\pm$  0.006 $\pm$   0.002 &  -0.004  $\pm$  0.005  $\pm$   0.014 &  0.001  $\pm$   0.005   $\pm$   0.004 \\
\hline
\multirow{6}{*}{\rotatebox{90}{\mbox{$-t [\text{GeV}^2]$}}} & 0.00-0.03 &  0.018 & 0.068 &  1.72 &  -0.014  $\pm$  0.009 $\pm$ 0.007 & 
0.006  $\pm$  0.012  $\pm$   0.003 &  -0.038  $\pm$  0.012 $\pm$  0.001 &  -0.022   $\pm$  0.012   $\pm$   0.004\\
& 0.03-0.06 &  0.043 & 0.088 &  2.26& -0.008  $\pm$  0.009  $\pm$   0.006 &
0.016 $\pm$  0.012  $\pm$   0.014 &  -0.004  $\pm$  0.012  $\pm$  0.007 &  0.003   $\pm$  0.012   $\pm$   0.005\\
& 0.06-0.10 &  0.078 & 0.099 &  2.51 & -0.025  $\pm$  0.009  $\pm$  0.007 & 
0.030 $\pm$  0.013  $\pm$   0.013 & 0.011  $\pm$  0.012 $\pm$   0.013 &  -0.028   $\pm$  0.012  $\pm$    0.004\\
& 0.10-0.20 &  0.142 & 0.110 &  2.79 &  -0.019  $\pm$  0.008   $\pm$  0.010 & 
0.036 $\pm$  0.012  $\pm$   0.014 &  0.007  $\pm$  0.011  $\pm$  0.025 & 0.019   $\pm$  0.011    $\pm$  0.001\\
& 0.20-0.35 &  0.260 & 0.121 &  3.27 &  -0.042 $\pm$   0.011  $\pm$  0.004 &
0.078 $\pm$  0.016  $\pm$ 0.029 & -0.016 $\pm$   0.015  $\pm$  0.040 & 0.023  $\pm$   0.015   $\pm$   0.001\\
& 0.35-0.70 &  0.460 & 0.125 &  3.82 &  -0.083  $\pm$  0.016  $\pm$   0.005 & 
0.069 $\pm$  0.025  $\pm$   0.061 & 0.052 $\pm$   0.022  $\pm$  0.040 & 0.030   $\pm$  0.021   $\pm$ 0.017\\
\hline
\multirow{6}{*}{\rotatebox{90}{\mbox{$x_{\text{B}}$}}} & 0.03-0.06 &  0.095 & 0.049 &  1.34 &  -0.038  $\pm$  0.009  $\pm$   0.013 & 
 0.007  $\pm$  0.013  $\pm$   0.013 & -0.026 $\pm$  0.012 $\pm$   0.013 &  -0.015   $\pm$  0.011  $\pm$    0.003\\
& 0.06-0.08 &  0.091 & 0.069 &  1.80&   -0.046  $\pm$  0.008  $\pm$   0.012 &
0.025  $\pm$  0.012  $\pm$   0.013 & -0.001  $\pm$ 0.011  $\pm$   0.009 & -0.003   $\pm$  0.011   $\pm$   0.005\\
& 0.08-0.10 &  0.104 & 0.089 &  2.30 &  -0.004  $\pm$  0.010  $\pm$   0.016 & 
0.033  $\pm$  0.014  $\pm$   0.019 & -0.013 $\pm$  0.013 $\pm$    0.017 & -0.013   $\pm$  0.013    $\pm$  0.002\\
& 0.10-0.13 &  0.121 &  0.113 &  2.93 &  -0.025  $\pm$  0.010  $\pm$   0.015 & 
0.035  $\pm$  0.014 $\pm$   0.002 & -0.015 $\pm$  0.013  $\pm$   0.006 & 0.001   $\pm$  0.013  $\pm$    0.009\\
& 0.13-0.20 &  0.159 & 0.157 &  4.06&   -0.012   $\pm$ 0.011  $\pm$   0.007 & 
0.029  $\pm$  0.015 $\pm$    0.031 & -0.003  $\pm$  0.014  $\pm$   0.001 & 0.016   $\pm$  0.014   $\pm$  0.005\\
& 0.20-0.35 &  0.231 & 0.244 &  6.14 &  -0.009 $\pm$  0.018   $\pm$  0.026 & 
0.060  $\pm$  0.026   $\pm$    0.018 & 0.069  $\pm$  0.024  $\pm$ 0.027 & 0.043  $\pm$   0.024  $\pm$   0.009\\
\hline
\multirow{6}{*}{\rotatebox{90}{\mbox{$Q^2 [\text{GeV}^2]$}}} & 1.00-1.40 &  0.076 & 0.054  & 1.20 &  -0.046  $\pm$  0.008  $\pm$   0.023 & 
0.021  $\pm$  0.012  $\pm$   0.028 &  -0.020 $\pm$  0.011  $\pm$  0.012 & -0.014  $\pm$  0.011   $\pm$   0.004\\
& 1.40-1.80 &  0.089 & 0.069 &  1.59 &  -0.018  $\pm$  0.009  $\pm$   0.021 & 
0.032  $\pm$  0.013  $\pm$   0.020 & -0.011  $\pm$  0.012  $\pm$  0.012 & 0.001  $\pm$  0.012   $\pm$  0.002\\
& 1.80-2.40 &  0.104 & 0.085 &  2.08 &  -0.039  $\pm$  0.009  $\pm$   0.022 &
0.016  $\pm$  0.012  $\pm$   0.017 & -0.008 $\pm$   0.012  $\pm$  0.013 & 0.018  $\pm$   0.012  $\pm$  0.004\\
& 2.40-3.20 &  0.126 & 0.105  & 2.77 &  -0.016 $\pm$   0.010  $\pm$   0.019 &  
0.068  $\pm$  0.014  $\pm$   0.020 & 0.006  $\pm$  0.013  $\pm$  0.013 & -0.022  $\pm$  0.013  $\pm$  0.009\\
& 3.20-4.50 &  0.151 & 0.134 &  3.76 &  0.002  $\pm$  0.011   $\pm$  0.014 & 
0.010 $\pm$   0.015  $\pm$   0.002 & -0.008  $\pm$  0.014 $\pm$ 0.015 & 0.014   $\pm$  0.014  $\pm$  0.002\\
& 4.50-10.0 &  0.218 & 0.200 &  5.82 &  -0.003  $\pm$  0.013  $\pm$   0.014 & 
0.037  $\pm$  0.018  $\pm$  0.038 & 0.042 $\pm$   0.017  $\pm$  0.003 & 0.010   $\pm$  0.017   $\pm$   0.005\\
\hline
  \end{tabular}
}
\caption{Results of the $A_{\textrm{C}}^{\cos(n\phi)}$ asymmetry
  amplitudes with statistical and systematic uncertainties and average
  values of the kinematic variables from unpolarised hydrogen target data taken during 2006-2007 at H{\sc ermes} for each $-t$, $x_{\textrm{B}}$ and $Q^{2}$ bin.}
\end{table}

\begin{table}[width=15cm]
\resizebox{16cm}{!} {
\begin{tabular}{|c|c|c|c|c|r|r|r|} \hline
\multicolumn{2}{|c|}{} & $\langle -t\rangle$ & $\langle
x_{\text{B}}\rangle$ & $\langle Q^2 \rangle $ & 
\multicolumn{1}{c|}{\multirow{2}{*}{$A_{\text{LU,I}}^{\sin \phi} \pm \delta_{stat.} \pm \delta_{syst.}$}} & 
\multicolumn{1}{c|}{\multirow{2}{*}{$A_{\text{LU,DVCS}}^{\sin \phi} \pm \delta_{stat.} \pm \delta_{syst.}$ }} & 
\multicolumn{1}{c|}{\multirow{2}{*}{$A_{\text{LU,I}}^{\sin (2\phi)} \pm \delta_{stat.} \pm \delta_{syst.}$}} \\ 
\multicolumn{2}{|c|}{} &  $[\text{GeV}^2]$ & & $[\text{GeV}^2]$ & & &  \\
\hline \hline
\multicolumn{2}{|c|}{overall} &  0.118 & 0.097 &  2.51 &  -0.229  $\pm$  0.018  $\pm$   0.024 &
 0.017  $\pm$  0.018  $\pm$  0.001 & -0.010  $\pm$  0.018  $\pm$   0.001 \\
\hline
\multirow{6}{*}{\rotatebox{90}{\mbox{$-t [\text{GeV}^2]$}}} & 0.00-0.03 &  0.019 & 0.069 &  1.72 &  -0.225  $\pm$  0.039 $\pm$   0.010 &
 0.048  $\pm$  0.039   $\pm$  0.003 & 0.003  $\pm$  0.039  $\pm$   0.003\\
& 0.03-0.06 &  0.044 & 0.088 &  2.25 &  -0.242 $\pm$   0.039   $\pm$  0.014 &
 0.019 $\pm$   0.039  $\pm$   0.005 & 0.026  $\pm$  0.038  $\pm$   0.001\\
& 0.06-0.10 &  0.079 & 0.099 &  2.49 & -0.177 $\pm$   0.043   $\pm$  0.012 &
 -0.023  $\pm$  0.043  $\pm$   0.004 & -0.002  $\pm$  0.043   $\pm$  0.005 \\
& 0.10-0.20 &  0.143 & 0.109 &  2.76 &  -0.253 $\pm$   0.037  $\pm$   0.010 &
0.010  $\pm$  0.037  $\pm$   0.002 & -0.008  $\pm$  0.037  $\pm$  0.006\\
& 0.20-0.35 &  0.261 & 0.119 &  3.23 &  -0.287 $\pm$   0.050  $\pm$   0.006 &
0.105  $\pm$  0.051  $\pm$   0.002 & -0.047 $\pm$   0.051   $\pm$  0.005\\
& 0.35-0.70 &  0.463 & 0.122 &  3.73 &  -0.173  $\pm$  0.072  $\pm$   0.016 & 
-0.107  $\pm$  0.073   $\pm$  0.023 & -0.111 $\pm$   0.074  $\pm$   0.028\\
\hline
\multirow{6}{*}{\rotatebox{90}{\mbox{$x_{\text{B}}$}}} & 0.03-0.06 &  0.099 & 0.049 & 1.34 & -0.249  $\pm$  0.038  $\pm$   0.079 &
0.035 $\pm$   0.038  $\pm$   0.004 & -0.044  $\pm$  0.039  $\pm$  0.011 \\ 
& 0.06-0.08 &  0.093 & 0.070 &  1.79 &  -0.228 $\pm$  0.037  $\pm$   0.049 &
0.013  $\pm$  0.038  $\pm$   0.002 & 0.023 $\pm$   0.037  $\pm$   0.012\\
& 0.08-0.10 &  0.106 & 0.089 &  2.30 &  -0.250 $\pm$   0.043  $\pm$   0.033 &
0.047 $\pm$   0.043  $\pm$   0.004 & 0.032  $\pm$  0.043  $\pm$   0.007\\
& 0.10-0.13 &  0.122 &  0.114 &  2.94 &  -0.237 $\pm$   0.045  $\pm$   0.030 &
-0.003  $\pm$  0.045 $\pm$    0.006 & -0.027 $\pm$ 0.045  $\pm$   0.010\\
& 0.13-0.20 &  0.160 & 0.157 &  4.06 &  -0.224 $\pm$   0.048  $\pm$   0.013 &
0.033  $\pm$  0.048 $\pm$    0.006 & -0.050 $\pm$   0.047  $\pm$   0.006\\
& 0.20-0.35 &  0.233 & 0.244 &  6.13 &  -0.077  $\pm$  0.085 $\pm$    0.022 &
-0.141  $\pm$  0.085  $\pm$   0.012 & -0.023 $\pm$  0.080 $ \pm$  0.007\\
\hline
\multirow{6}{*}{\rotatebox{90}{\mbox{$Q^2 [\text{GeV}^2]$}}} & 1.00-1.40 &  0.078 & 0.055  & 1.20  &  -0.218  $\pm$  0.038  $\pm$   0.063 &
0.029 $\pm$   0.038  $\pm$   0.002 & -0.023  $\pm$  0.038  $\pm$   0.012\\
& 1.40-1.80 &  0.092 & 0.069 &  1.59  &  -0.257  $\pm$  0.042  $\pm$   0.064 &
0.040 $\pm$  0.042  $\pm$   0.006 & 0.011 $\pm$   0.042  $\pm$   0.004\\
& 1.80-2.40 &  0.106 & 0.085 &  2.08  &  -0.233 $\pm$  0.041  $\pm$   0.042 &
0.016 $\pm$   0.041  $\pm$   0.004 & -0.010  $\pm$  0.040  $\pm$   0.007\\
& 2.40-3.20 &  0.127 & 0.105  & 2.77  &  -0.302  $\pm$  0.044   $\pm$  0.044 & 
0.087  $\pm$  0.044  $\pm$   0.006 & 0.031  $\pm$  0.044  $\pm$   0.005\\
& 3.20-4.50 &  0.152 & 0.134 &  3.77  &  -0.160  $\pm$  0.048  $\pm$   0.032 &
-0.057  $\pm$  0.048  $\pm$   0.005 & -0.061  $\pm$  0.048  $\pm$   0.004 \\
& 4.50-10.0 &  0.220 & 0.199 &  5.79  &  -0.169 $\pm$ 0.057  $\pm$   0.008 &
-0.065  $\pm$  0.057 $\pm$ 0.005 & -0.017  $\pm$  0.056  $\pm$  0.013\\
\hline
  \end{tabular}
}
\caption{Results of the $A_{\textrm{LU,I}}^{\sin(n\phi)}$ and
  $A_{\textrm{LU,DVCS}}^{\sin \phi}$ asymmetry amplitudes with
  statistical and systematic uncertainties and the average values of
  the kinematic variables from unpolarised hydrogen target data taken during 1996-2007 at H{\sc ermes} for each $-t$, $x_{\textrm{B}}$ and $Q^{2}$ bin.
An additional 3.2\,\% scale uncertainty is present in the amplitudes due to the uncertainty of the beam polarisation measurement.
}
\end{table}

\begin{table}[width=15cm]
\resizebox{16cm}{!} {
\begin{tabular}{|c|c|c|c|c|r|r|r|r|} \hline
\multicolumn{2}{|c|}{} & $\langle -t\rangle$ & $\langle
x_{\text{B}}\rangle$ & $\langle Q^2 \rangle $ & 
\multicolumn{1}{c|}{\multirow{2}{*}{$A_{\text{C}}^{\cos (0\phi)} \pm \delta_{stat.} \pm \delta_{syst.}$ }} & 
\multicolumn{1}{c|}{\multirow{2}{*}{$A_{\text{C}}^{\cos \phi } \pm \delta_{stat.} \pm \delta_{syst.}$}} & 
\multicolumn{1}{c|}{\multirow{2}{*}{$A_{\text{C}}^{\cos (2\phi) }\pm \delta_{stat.} \pm \delta_{syst.}$}} &
\multicolumn{1}{c|}{\multirow{2}{*}{$A_{\text{C}}^{\cos (3\phi) } \pm \delta_{stat.} \pm \delta_{syst.}$}} \\ 
\multicolumn{2}{|c|}{} &  $[\text{GeV}^2]$ & & $[\text{GeV}^2]$ & &  &  &  \\
\hline
\hline
\multicolumn{2}{|c|}{overall} &  0.119 & 0.097 &  2.51 &  -0.021 $\pm$  0.003 $\pm$  0.010 & 
0.041  $\pm$  0.005 $\pm$   0.002 &  -0.003  $\pm$  0.005  $\pm$   0.014 &  -0.002  $\pm$   0.005   $\pm$   0.003 \\
\hline
\multirow{6}{*}{\rotatebox{90}{\mbox{$-t [\text{GeV}^2]$}}} & 0.00-0.03 &  0.019 & 0.069 & 1.72  &  -0.017  $\pm$  0.007 $\pm$ 0.007 & 
0.005  $\pm$  0.010  $\pm$   0.003 &  -0.023  $\pm$  0.010 $\pm$  0.001 &  -0.013   $\pm$  0.010   $\pm$   0.004\\
& 0.03-0.06 &  0.044 & 0.088 & 2.25 & -0.005  $\pm$  0.007  $\pm$   0.006 &
0.007 $\pm$  0.010  $\pm$   0.014 &  -0.003  $\pm$  0.010  $\pm$  0.007 &  0.005   $\pm$  0.010   $\pm$   0.004\\
& 0.06-0.10 & 0.079  & 0.099 &  2.49 & -0.012  $\pm$  0.008  $\pm$  0.006 & 
0.028 $\pm$  0.011  $\pm$   0.013 & 0.013  $\pm$  0.011 $\pm$   0.013 &  -0.023   $\pm$  0.011  $\pm$    0.003\\
& 0.10-0.20 & 0.143  & 0.109 &  2.76 &  -0.016  $\pm$  0.007   $\pm$  0.009 & 
0.052 $\pm$  0.009  $\pm$   0.015 &  -0.008  $\pm$  0.009  $\pm$  0.025 & 0.006   $\pm$  0.009    $\pm$  0.001\\
& 0.20-0.35 &   0.261 & 0.119 &  3.23 &  -0.040 $\pm$   0.009  $\pm$  0.002 &
0.108 $\pm$  0.013  $\pm$ 0.030 & -0.003 $\pm$   0.013  $\pm$  0.040 & 0.012  $\pm$   0.013   $\pm$   0.001\\
& 0.35-0.70 &  0.462 & 0.122 &  3.73 &  -0.072  $\pm$  0.014  $\pm$   0.004 & 
0.134 $\pm$  0.021  $\pm$   0.062 & 0.049 $\pm$   0.019  $\pm$  0.040 & 0.030   $\pm$  0.019   $\pm$ 0.017\\
\hline
\multirow{6}{*}{\rotatebox{90}{\mbox{$x_{\text{B}}$}}} & 0.03-0.06 &  0.099 &  0.049 &   1.34 &  -0.045  $\pm$  0.007  $\pm$   0.014 & 
0.016  $\pm$  0.011  $\pm$   0.014 & -0.017 $\pm$  0.010 $\pm$  0.013 &  0.004   $\pm$  0.009  $\pm$    0.003\\
& 0.06-0.08 & 0.093  & 0.070 & 1.79  &   -0.035  $\pm$  0.007  $\pm$   0.013 &
0.028  $\pm$  0.009  $\pm$   0.015 & -0.019  $\pm$ 0.009  $\pm$  0.009 & -0.012   $\pm$  0.009   $\pm$   0.005\\
& 0.08-0.10 &  0.106 & 0.089 &  2.30 &  -0.017  $\pm$  0.008  $\pm$   0.017 & 
0.044  $\pm$  0.011  $\pm$   0.019 & 0.005 $\pm$  0.011 $\pm$    0.016 & -0.009   $\pm$  0.011    $\pm$  0.002\\
& 0.10-0.13 &  0.122 & 0.114  & 2.94  &  -0.007  $\pm$  0.008  $\pm$   0.015 & 
0.030  $\pm$  0.012 $\pm$   0.002 & -0.001 $\pm$  0.012  $\pm$   0.006 & -0.009   $\pm$  0.011  $\pm$    0.009\\
& 0.13-0.20 &  0.160 & 0.157 & 4.06 &   -0.006   $\pm$ 0.009  $\pm$   0.007 & 
0.049  $\pm$  0.013 $\pm$    0.030 & -0.001  $\pm$  0.012  $\pm$   0.001 & 0.002   $\pm$  0.012   $\pm$  0.005\\
& 0.20-0.35 & 0.233  & 0.244 &  6.13 &  0.019 $\pm$  0.016   $\pm$  0.027 & 
0.050  $\pm$  0.022   $\pm$  0.018 & 0.051  $\pm$  0.022  $\pm$   0.027 & 0.006  $\pm$   0.021  $\pm$   0.008\\
\hline
\multirow{6}{*}{\rotatebox{90}{\mbox{$Q^2 [\text{GeV}^2]$}}} & 1.00-1.40 &  0.078 &  0.055 & 1.20 &  -0.048  $\pm$  0.007  $\pm$   0.024 & 
0.029  $\pm$  0.010  $\pm$   0.030 &  -0.008 $\pm$  0.009  $\pm$  0.011 & 0.005  $\pm$  0.009   $\pm$   0.004\\
& 1.40-1.80 & 0.092  & 0.069 &  1.59 &  -0.022  $\pm$  0.008  $\pm$   0.022 & 
0.041  $\pm$  0.011  $\pm$   0.021 & -0.020  $\pm$  0.011  $\pm$  0.012 & -0.004  $\pm$  0.011   $\pm$  0.002\\
& 1.80-2.40 &  0.106 & 0.085 &  2.08 &  -0.028  $\pm$  0.007  $\pm$   0.022 &
 0.031  $\pm$  0.010  $\pm$   0.018 & -0.012 $\pm$   0.010  $\pm$  0.013 & -0.000  $\pm$   0.010  $\pm$  0.004\\
& 2.40-3.20 &  0.127 &  0.105 & 2.77 &  -0.017 $\pm$   0.008  $\pm$   0.019 &  
0.059  $\pm$  0.012  $\pm$   0.019 & 0.014  $\pm$  0.011  $\pm$  0.013 & -0.010  $\pm$  0.011  $\pm$  0.009\\
& 3.20-4.50 &   0.152 & 0.134 &  3.77 &  0.011  $\pm$  0.009   $\pm$  0.013 & 
0.037 $\pm$   0.013  $\pm$   0.002 & -0.008  $\pm$  0.013 $\pm$ 0.015 & -0.012   $\pm$  0.012  $\pm$  0.001\\
& 4.50-10.0 & 0.220  & 0.199 & 5.79  &  0.006  $\pm$  0.011  $\pm$   0.013 & 
0.038  $\pm$  0.015  $\pm$  0.038 & 0.036 $\pm$   0.015  $\pm$  0.002 & 0.005   $\pm$  0.015   $\pm$   0.004\\
\hline
  \end{tabular}
}
\caption{Results of the $A_{\textrm{C}}^{\cos(n\phi)}$ asymmetry
  amplitudes with statistical and systematic uncertainties and the
  average values of the
  kinematic variables from unpolarised hydrogen target data taken during 1996-2007 at H{\sc ermes} for each $-t$, $x_{\textrm{B}}$ and $Q^{2}$ bin.}
\end{table}

\begin{table}[width=15cm]
\resizebox{16cm}{!} {
\begin{tabular}{|cc|c|c|c|c|r|r|r|} \hline
\multicolumn{3}{|c|}{} & $\langle -t\rangle$ & $\langle
x_{\text{B}}\rangle$ & $\langle Q^2 \rangle $ & 
\multicolumn{1}{c|}{\multirow{2}{*}{$A_{\text{LU,I}}^{\sin \phi} \pm \delta_{stat.} \pm \delta_{syst.}$ }} & 
\multicolumn{1}{c|}{\multirow{2}{*}{$A_{\text{LU,DVCS}}^{\sin \phi } \pm \delta_{stat.} \pm \delta_{syst.}$}} & 
\multicolumn{1}{c|}{\multirow{2}{*}{$A_{\text{LU,I}}^{\sin (2\phi)} \pm \delta_{stat.} \pm \delta_{syst.}$}} \\ 
\multicolumn{3}{|c|}{} &  $[\text{GeV}^2]$ & & $[\text{GeV}^2]$ & & &  \\
\hline \hline
\multirow{6}{*}{\rotatebox{90}{\mbox{$-t [\text{GeV}^2]$}}} & \multirow{6}{*}{\rotatebox{90}{\mbox{$ 0.03 < x_{\text{B}} < 0.08$}}} & 0.00-0.03 &  0.018 & 0.058  & 1.473  &  -0.253  $\pm$  0.046  $\pm$  0.019  &
 0.092  $\pm$   0.046  $\pm$ 0.008  &   -0.039 $\pm$  0.046  $\pm$  0.014 \\
& & 0.03-0.06 & 0.043  &  0.060 &  1.558 &  -0.188 $\pm$  0.058    $\pm$  0.024 &
 -0.021 $\pm$ 0.058  $\pm$  0.007 &  0.074 $\pm$  0.057  $\pm$  0.007 \\
& & 0.06-0.10 &  0.078 & 0.060 &  1.567 & -0.241  $\pm$  0.069 $\pm$  0.018 &
  -0.013 $\pm$  0.069  $\pm$  0.004  &  -0.034 $\pm$  0.068   $\pm$   0.012\\
& & 0.10-0.20 &  0.142 & 0.060 & 1.576  &  -0.261 $\pm$  0.062   $\pm$  0.020  &
 -0.015 $\pm$  0.062  $\pm$  0.006  &  0.025 $\pm$  0.062  $\pm$ 0.009 \\
& & 0.20-0.35 &  0.259 & 0.057 & 1.701  & -0.284  $\pm$ 0.087   $\pm$  0.014 &
 0.145 $\pm$  0.088  $\pm$   0.006 &  -0.041 $\pm$ 0.091  $\pm$ 0.010 \\
& & 0.35-0.70 & 0.465  &  0.054 &  1.819 &  -0.126  $\pm$  0.147 $\pm$ 0.008  & 
 -0.120 $\pm$  0.147   $\pm$ 0.053  &  -0.115  $\pm$   0.159  $\pm$ 0.016 \\
\hline
\multirow{6}{*}{\rotatebox{90}{\mbox{$-t [\text{GeV}^2]$}}} & \multirow{6}{*}{\rotatebox{90}{\mbox{$ 0.08 < x_{\text{B}} < 0.12$ }}} & 0.00-0.03 &  0.022  &0.095  & 2.311  &  -0.163  $\pm$   0.078 $\pm$   0.008 &
 -0.063 $\pm$   0.079  $\pm$   0.003 &  0.096  $\pm$  0.079  $\pm$  0.016 \\
& & 0.03-0.06 &  0.044 & 0.098 &  2.501 &  -0.302  $\pm$ 0.069   $\pm$  0.008  &
 0.053 $\pm$  0.070  $\pm$   0.007 & -0.024 $\pm$   0.068  $\pm$ 0.002  \\
& & 0.06-0.10 & 0.079  & 0.098 & 2.462  &  -0.187 $\pm$  0.078  $\pm$  0.007  &
 0.003 $\pm$  0.078   $\pm$  0.007 &  -0.049 $\pm$  0.078  $\pm$  0.007  \\
& & 0.10-0.20 & 0.142  &  0.098 & 2.484  &  -0.237  $\pm$   0.068  $\pm$ 0.005  & 
 0.009 $\pm$   0.068 $\pm$   0.010 &  -0.053 $\pm$  0.068  $\pm$  0.019 \\
& & 0.20-0.35 &  0.258 & 0.099 & 2.736  &   -0.280 $\pm$  0.099  $\pm$  0.003  &
 0.077 $\pm$   0.100 $\pm$  0.005 &  -0.031  $\pm$  0.101  $\pm$  0.023  \\
& & 0.35-0.70 &  0.459 & 0.099 & 3.211  &  -0.278 $\pm$ 0.151  $\pm$ 0.010   &
 -0.025 $\pm$  0.152 $\pm$  0.014  &  0.150 $\pm$  0.163  $\pm$ 0.007 \\
\hline
\multirow{6}{*}{\rotatebox{90}{\mbox{$-t [\text{GeV}^2]$}}} & \multirow{6}{*}{\rotatebox{90}{\mbox{$ 0.12 < x_{\text{B}} < 0.35$}}} & 0.00-0.03 & 0.026  & 0.130  & 2.954 &  -0.238  $\pm$  0.238 $\pm$ 0.066  &
0.006 $\pm$  0.237  $\pm$  0.052 &  0.275 $\pm$  0.239 $\pm$ 0.041 \\
& & 0.03-0.06 & 0.046  & 0.145 & 3.629  &  -0.235 $\pm$ 0.091   $\pm$  0.009  &
 0.037 $\pm$  0.092  $\pm$ 0.008  & 0.006 $\pm$  0.094 $\pm$ 0.009 \\
& & 0.06-0.10 & 0.080  & 0.160 & 3.942  &  -0.086 $\pm$ 0.085  $\pm$  0.011 &
-0.031 $\pm$   0.085 $\pm$  0.003  &  0.058 $\pm$ 0.083  $\pm$  0.007\\
& & 0.10-0.20 & 0.145  &  0.174 & 4.309 &  -0.248  $\pm$  0.067  $\pm$ 0.004   &
  0.029 $\pm$ 0.068  $\pm$  0.002  &  -0.004  $\pm$  0.066 $\pm$  0.007 \\
& & 0.20-0.35 & 0.263  & 0.184 &  4.799 &  -0.283 $\pm$  0.085  $\pm$ 0.025  &
 0.133 $\pm$  0.085 $\pm$   0.010 &  -0.061  $\pm$  0.082   $\pm$  0.005\\
& & 0.35-0.70 & 0.460  & 0.194 & 5.621  &  -0.117  $\pm$  0.117  $\pm$  0.046 &
 -0.200 $\pm$  0.117  $\pm$  0.024  & -0.218 $\pm$ 0.110 $ \pm$ 0.048 \\
\hline
  \end{tabular}
}
\caption{Results of the $A_{\textrm{LU,I}}^{\sin(n\phi)}$ and
  $A_{\textrm{LU,DVCS}}^{\sin \phi}$ asymmetry amplitudes with
  statistical and systematic uncertainties and the average values of
  the kinematic variables from unpolarised hydrogen target data taken during 1996-2007 at H{\sc ermes} for $-t$ bins with certain $x_{\textrm{B}}$ ranges.
An additional 3.2\,\% scale uncertainty is present in the amplitudes due to the uncertainty of
the beam polarisation measurement.}
\end{table}

\begin{table}[width=15cm]
\resizebox{16cm}{!} {
\begin{tabular}{|cc|c|c|c|c|r|r|r|r|} \hline
\multicolumn{3}{|c|}{} & $\langle -t\rangle$ & $\langle
x_{\text{B}}\rangle$ & $\langle Q^2 \rangle $ & 
\multicolumn{1}{c|}{\multirow{2}{*}{$A_{\text{C}}^{\cos (0\phi)} \pm \delta_{stat.} \pm \delta_{syst.}$}} & 
\multicolumn{1}{c|}{\multirow{2}{*}{$A_{\text{C}}^{\cos \phi } \pm \delta_{stat.} \pm \delta_{syst.}$}}& 
\multicolumn{1}{c|}{\multirow{2}{*}{$A_{\text{C}}^{\cos (2\phi) } \pm \delta_{stat.} \pm \delta_{syst.}$}}&
\multicolumn{1}{c|}{\multirow{2}{*}{$A_{\text{C}}^{\cos (3\phi) } \pm \delta_{stat.} \pm \delta_{syst.}$}} \\ 
\multicolumn{3}{|c|}{} &  $[\text{GeV}^2]$ & & $[\text{GeV}^2]$ & & & & \\
\hline
\hline
\multirow{6}{*}{\rotatebox{90}{\mbox{$-t [\text{GeV}^2]$}}} & \multirow{6}{*}{\rotatebox{90}{\mbox{$0.03 < x_{\text{B}} < 0.08$}}} & 0.00-0.03 &  0.018 & 0.058  & 1.473 &  -0.031  $\pm$  0.008 $\pm$ 0.006 & 
-0.013  $\pm$ 0.016  $\pm$ 0.002  &  -0.032 $\pm$  0.012 $\pm$ 0.004 &   -0.016  $\pm$   0.012 $\pm$ 0.003  \\
& & 0.03-0.06 & 0.043  &  0.060 &  1.558 &  -0.007 $\pm$  0.015 $\pm$ 0.008  &
0.019 $\pm$  0.015  $\pm$ 0.012  &  -0.019  $\pm$  0.015 $\pm$ 0.010 &   0.013 $\pm$  0.015  $\pm$  0.005 \\
& & 0.06-0.10 &  0.078 & 0.060 &  1.567 & -0.025  $\pm$  0.012 $\pm$ 0.017  & 
0.007 $\pm$ 0.017   $\pm$ 0.004  &  0.010 $\pm$ 0.017  $\pm$ 0.018  &  -0.014  $\pm$  0.017  $\pm$ 0.007 \\
& & 0.10-0.20 &  0.142 & 0.060 & 1.576 &  -0.041 $\pm$   0.011 $\pm$ 0.019  & 
 0.036 $\pm$ 0.017  $\pm$   0.001 &  -0.009  $\pm$ 0.016  $\pm$ 0.026 & 0.008  $\pm$  0.015   $\pm$ 0.002 \\
& & 0.20-0.35 &  0.259 & 0.057 & 1.701 &  -0.039 $\pm$  0.026  $\pm$ 0.023  &
0.148 $\pm$  0.044 $\pm$  0.003 & 0.021 $\pm$   0.036 $\pm$ 0.038 & 0.061  $\pm$ 0.027   $\pm$  0.017 \\
& & 0.35-0.70 & 0.465  &  0.054 &  1.819 &  -0.234  $\pm$  0.090  $\pm$  0.026  & 
-0.054 $\pm$ 0.158 $\pm$ 0.062  &  -0.109 $\pm$  0.113 $\pm$  0.064 &  -0.084  $\pm$  0.061  $\pm$ 0.020 \\
\hline
\multirow{6}{*}{\rotatebox{90}{\mbox{$-t [\text{GeV}^2]$}}} & \multirow{6}{*}{\rotatebox{90}{\mbox{$0.08 < x_{\text{B}} < 0.12$}}} & 0.00-0.03 &  0.022  &0.095  & 2.311 &  0.020  $\pm$ 0.015   $\pm$  0.014  & 
 0.049 $\pm$  0.021 $\pm$ 0.017  & -0.002 $\pm$ 0.021  $\pm$  0.008 &  -0.014  $\pm$ 0.021   $\pm$ 0.005 \\
& & 0.03-0.06 &  0.044 & 0.098 &  2.501  &  -0.013  $\pm$  0.013  $\pm$  0.014  &
-0.002  $\pm$  0.018 $\pm$ 0.003 & 0.037  $\pm$  0.018 $\pm$ 0.011  & 0.000  $\pm$ 0.018   $\pm$  0.003\\
& & 0.06-0.10 & 0.079  & 0.098 & 2.462 &  0.007  $\pm$  0.014  $\pm$  0.010  & 
 0.037 $\pm$ 0.020 $\pm$  0.002 & 0.012 $\pm$ 0.020 $\pm$ 0.010  &  -0.019 $\pm$  0.020  $\pm$ 0.010 \\
& & 0.10-0.20 & 0.142  & 0.098 & 2.484  &  -0.020 $\pm$  0.013 $\pm$   0.026 & 
0.038  $\pm$ 0.018 $\pm$ 0.018  & -0.024 $\pm$  0.017  $\pm$ 0.024 &  -0.005 $\pm$ 0.017  $\pm$ 0.007 \\
& & 0.20-0.35 &  0.258 & 0.099 & 2.736 &   -0.064  $\pm$ 0.018  $\pm$   0.029 & 
0.107  $\pm$ 0.025 $\pm$ 0.008 &  -0.011 $\pm$  0.024  $\pm$  0.042 &  -0.038 $\pm$ 0.024  $\pm$ 0.013 \\
& & 0.35-0.70 &  0.459 & 0.099 & 3.211 &  -0.077 $\pm$   0.038 $\pm$ 0.023  & 
0.023  $\pm$  0.063  $\pm$ 0.015 &  -0.046 $\pm$ 0.056  $\pm$  0.010 & 0.024 $\pm$ 0.048 $\pm$ 0.029 \\
\hline
\multirow{6}{*}{\rotatebox{90}{\mbox{$-t [\text{GeV}^2]$}}} & \multirow{6}{*}{\rotatebox{90}{\mbox{$0.12 < x_{\text{B}} < 0.35$}}} & 0.00-0.03 & 0.026  & 0.130  & 2.954 &  -0.024 $\pm$  0.045  $\pm$   0.008 & 
 -0.007 $\pm$ 0.064  $\pm$ 0.001  & -0.013 $\pm$ 0.065 $\pm$ 0.023 & 0.026  $\pm$ 0.062 $\pm$ 0.011 \\
& & 0.03-0.06 & 0.046  & 0.145 & 3.629 &  0.034  $\pm$   0.018 $\pm$   0.019 & 
 -0.044 $\pm$ 0.027  $\pm$ 0.045  & -0.005  $\pm$ 0.025 $\pm$ 0.020 & -0.027 $\pm$ 0.025 $\pm$ 0.014 \\
& & 0.06-0.10 & 0.080  & 0.160 & 3.942 &  -0.026 $\pm$   0.016 $\pm$  0.013  &
 0.044 $\pm$ 0.022  $\pm$ 0.038  & 0.012 $\pm$  0.022  $\pm$ 0.003  & -0.043 $\pm$ 0.021 $\pm$ 0.005 \\
& & 0.10-0.20 & 0.145  &  0.174 & 4.309 &  0.013 $\pm$    0.012 $\pm$   0.010 &  
 0.059 $\pm$  0.017 $\pm$ 0.024  & 0.002  $\pm$ 0.017   $\pm$ 0.010  &  0.011 $\pm$  0.017  $\pm$ 0.003 \\
& & 0.20-0.35 & 0.263  & 0.184 &  4.799 &  -0.000  $\pm$ 0.015   $\pm$  0.013 & 
0.069  $\pm$ 0.022  $\pm$  0.034 &  0.017 $\pm$ 0.021  $\pm$  0.023 &  0.010  $\pm$  0.021 $\pm$ 0.005 \\
& & 0.35-0.70 & 0.460  & 0.194 & 5.621  &   -0.023 $\pm$   0.021 $\pm$ 0.022  & 
 0.078 $\pm$  0.030 $\pm$  0.009 & 0.068 $\pm$   0.030  $\pm$ 0.020 &  0.061 $\pm$  0.029  $\pm$ 0.024 \\
\hline
  \end{tabular}
}
\caption{Results of the $A_{\textrm{C}}^{\cos(n\phi)}$ asymmetry
  amplitudes with statistical and systematic uncertainties and the
  average of the
  kinematic variables from unpolarised hydrogen target data taken during the 1996-2007 at H{\sc ermes} for $-t$ bins with certain $x_{\textrm{B}}$ ranges.}
\end{table}
\clearpage

\section{Covariance matrix results}

\begin{figure}[tbh]
\centering
\includegraphics[width=0.95\textwidth]{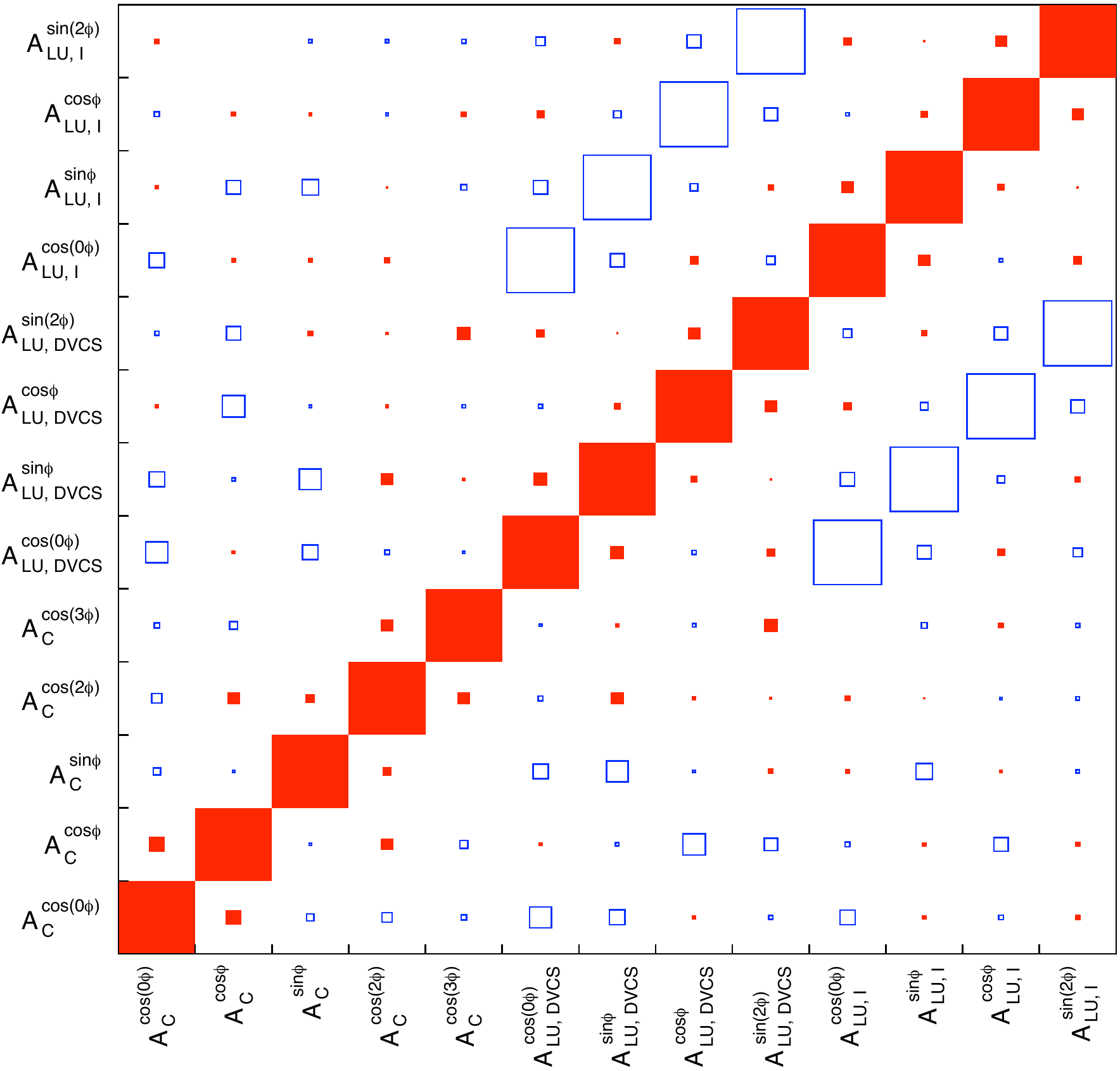}
\caption{The covariance matrix results for the asymmetries extracted in a single 
bin across the whole kinematic range. The size of the symbols in
the chart reflect the magnitude of the corresponding correlation. Closed
(open) symbols represent positive (negative) correlations.}
\label{pic:covmat}
\end{figure}

Figure~\ref{pic:covmat} shows the covariance matrix for the 13 parameter fit for all amplitudes extracted in a single bin across the entire kinematic range. The covariance matrix for the fit in each of the kinematic bins presented in figures~\ref{bsa_xbjrange} to~\ref{bca_xbjrange2} will be made available in the Durham database.

\bibliographystyle{natbib}

\end{document}